\documentclass[10pt,english,aps,showkeys,notitlepage]{revtex4-1}
\pdfoutput=1
\usepackage[T1]{fontenc}
\usepackage{color}
\usepackage{amsmath}
\usepackage{amssymb}
\usepackage{graphicx}
\usepackage{esint}

\usepackage{babel}
\begin{document}

%%%%%%%%%%%%%%%% FRONT PAGE %%%%%%%%%%%%%%%%

\title{Circular Coloring of Random Graphs: Statistical Physics
  Investigation}

\author{Christian Schmidt$^{1}$, Nils-Eric Guenther$^{1,2}$, and Lenka Zdeborov\'a$^{1}$}

\affiliation{
$^1$ Institut de Physique Th\'eorique, CEA Saclay and CNRS, 91191, Gif-sur-Yvette, France.
$^2$ ICFO -- Institut de Ci\`encies Fot\`oniques, The Barcelona Institute of Science and Technology, 08860 Castelldefels, Spain.
}

\begin{abstract}
Circular coloring is a constraints satisfaction problem where colors
are assigned to nodes in a graph in such a way that every pair of
connected nodes has two consecutive colors (the first color being consecutive to
the last). We study circular coloring of random graphs using the
cavity method. We identify two very interesting properties of this
problem. For sufficiently many color and sufficiently low temperature
there is a spontaneous breaking of the circular symmetry between
colors and a phase transition forwards a ferromagnet-like phase. Our
second main result concerns 5-circular coloring of random 3-regular graphs. While this case is
found colorable, we conclude that the description via one-step replica
symmetry breaking is not sufficient. We observe that simulated
annealing is very efficient to find proper colorings for this
case. The 5-circular coloring of 3-regular random graphs thus provides a first known example of a problem
where the ground state energy is known to be exactly zero yet the
space of solutions probably requires a full-step replica symmetry
breaking treatment. 
\end{abstract}

\maketitle

%%%%%%%%%%%%%%%%%%%%%%%%%%%%%%%%%%%%%%%%

%%%%%%%%%%%%%%%% INTRO %%%%%%%%%%%%%%%%

\section{Introduction}

This paper inscribes in a line of work where
statistical physics methods such as the cavity method, developed in the field of spin glasses
\cite{MezardParisi87b,MezardMontanari07}, are applied to study random
instances of constraint satisfaction problems. The most well known
works in this direction are those of random graph coloring and
random $K$-satisfiability
\cite{MezardParisi02,MuletPagnani02,KrzakalaMontanari06,ZdeborovaKrzakala07}.

This paper treats the problem of {\it circular coloring} of random
graphs. Whereas in the canonical coloring two nodes of a graph that
are connected by an edge are required to have different colors, in
circular coloring the colors are ordered into a circle, and two
adjacent nodes are required to have two adjacent colors. 

To define circular coloring consider a graph $\mathcal{G}=(\mathcal{V},\mathcal{E})$
where each node $i\in\mathcal{V=}\{1,\dots,N\}$ can attain the discrete
values (colors) $s_{i}\in\{1,2,\dots,q\}$ and two nodes $(i,j)$
are connected if $(i,j)\in\mathcal{E}$. We denote $j$ as a neighbor
of $i$, $j\in N(i)$, if $(i,j)\in\mathcal{E}$. Then the graph is  $q$-circular colorable
if and only if there exists an assignment of $q$ colors to the nodes such that, if a node $i \in
\mathcal{V}$ is of color $s_i$, then all nodes $j \in N(i)$ are of
color $s_j \in \{ s_i-1,s_i+1\}$ modulo $q$. The smallest $q$ for
which a given graph is $q$-circular colorable is called the {\it
  circular chromatic number} of that graph. 

The main motivation
for the present work is the existence of a conjecture by Ne\v{s}et\v{r}il that states that all graphs of
maximum degree three (sub-cubic) without short cycles are 5-circular
colorable, cf. ``The Pentagon Problem'' in
\cite{nevsetvril2013combinatorial}. The results of the cavity method confirm that random graphs
of degree three are indeed 5-circular colorable. Next to this our
solution unveils several striking properties of this problem that have broader
interest, and we present them as the main results of this paper. 

Note that for $q=2$ and $q=3$ there is no difference between the
canonical and circular coloring. Hence all the works on canonical
coloring apply. For this reason the present paper will consider only
$q > 3$ (and mostly odd $q$, see below).

\subsection{Context in mathematics\label{sub:math}}

Circular coloring belongs to a larger class of problems that generalizes the
canonical graph coloring problem and is often explained using graph
homomorphisms, objects of more general interest in mathematics. Given graphs $\mathcal{G}=(\mathcal{V},\mathcal{E})$ and $\mathcal{G}'=(\mathcal{V}',\mathcal{E}')$, a {\it
  homomorphism} is any mapping $f: \mathcal{V} \to \mathcal{V}'$ which satisfies $(ij)\in
\mathcal{E} \Rightarrow (f(i)f(j)) \in \mathcal{E}'$. The existence of a coloring of a graph  $\mathcal{G}=(\mathcal{V},\mathcal{E})$ with
$q$ colors is hence equivalent to an existence of a homomorphism of
that graph onto a complete graph on $q$ nodes. Circular coloring is
equivalent to a homomorphism onto a cycle on $q$ nodes. Clearly, all the
other possibilities for the graph $\mathcal{G}'$ are of interest in
mathematics.

Ne\v{s}et\v{r}il's Pentagon Conjecture states that there exists an
integer $l$ such that if every node in a graph has degree at most 3,
and no cycles shorter than $l$ (i.e. so called girth at least $l$) then such a graph is 5-circular
colorable \cite{nevsetvril2013combinatorial}. This conjecture is
inspired by the aim to generalize classical results known for
coloring, for instance that every graph with maximum degree 3 is
3-colorable unless it contains a complete graph of 4 nodes
\cite{brooks1941colouring}.  

Note that circular coloring with an even number of colors is closely
related to 2-coloring. Indeed, if an optimal assignment with a given
set of violated edges exist for the 2-coloring, then the same set of violated 
edges is also optimal for $2q$-circular coloring for any integer $q$ (we simply use only 2
of the $2q$ colors). On the other hand if a graph is $2q$-circular
colorable with a given number of violated edges for some integer $q$
then it is also 2-colorable with less or equal number of violated edges because
every odd color can be replaced by the first one and every even color
by the second one.  Therefore ground state properties of $q>2$ circular coloring are only interesting
when the number of colors $q$ is odd. 

A series of mathematical works established that there are sub-cubic
graphs with large girth that are not $q$-circular colorable for $q\ge
7$ \cite{kostochka2001colorings,wanless2001regular,hatami2005random}. These proofs show that a random 3-regular graph is not
$q$-circular colorable for $q\ge 7$ using variants of the first moment
method. Therefore, $q=5$ is the
remaining open case for colorability of sub-cubic graphs. The use of random graphs in the
proofs \cite{wanless2001regular,hatami2005random} is an important motivation to study the behavior of 5-circular coloring on
the same class of graphs. Another motivation is the existence of powerful,
non-rigorous techniques, from statistical physics, that can easily be
adapted to study circular colorings on random graphs. 
Circular coloring of particular graphs or deterministic classes of
graphs is well studied, for a recent review and references see
\cite{zhu2006recent}. Concerning circular coloring on random graphs not much is
known, apart from \cite{wanless2001regular,hatami2005random}.

From the known mathematical results the most remarkable one is perhaps
the one of \cite{hatami2005random} that established the non-colorability
for 7-circular coloring of random 3-regular graphs. A simple
calculation of the expected number of proper circular colorings where
every color is present on the same number of nodes shows that the
vanilla 1st moment method is not sufficient to show that 3-regular
graphs are with high probability not 7-circular colorable. This
suggest that something non-trivial is happening for $7$-circular
coloring of random $3$ regular graphs. And the upper bound
established by \cite{hatami2005random} is non-trivial along the
lines of the upper bound of \cite{coja2013upper} for coloring. 

\subsection{Summary of our main results}

We apply the cavity method of spin glasses to circular coloring on
random (mostly regular) graphs in its replica symmetric and one-step
replica symmetry broken (1RSB) version
\cite{MezardParisi01,MezardParisi03}. Compared to all the other
combinatorial problems that were treated with these techniques
previously, circular coloring is special in the three following interesting
ways. 

\subsubsection{Spontaneous breaking of color symmetry}

Circular coloring enjoyes a global symmetry: colors in a proper coloring
can be reflected and shifted by a constant without violating any edges. In canonical
coloring any permutation of colors has this property. In canonical
coloring, the statistical physics works \cite{MuletPagnani02,KrzakalaMontanari06,ZdeborovaKrzakala07} always assume that
even if this global symmetry is lifted, every color appears roughly
on the same number of nodes. Configurations where some color is more
represented than others are thermodynamically subdominant. For
canonical coloring this assumption seems to be consistent within the
cavity method, but it is
not known how to prove it rigorously, as discussed in \cite{dani2012tight}.

Circular coloring with sufficiently large $q$ and temperature
sufficiently low is drastically
different. The cyclic symmetry between colors is broken and configurations
where a subset of colors is much more represented than others are the
thermodynamically dominant ones. We call the phase where such a
symmetry breaking appears the ``ferromagnetic'' phase. Interestingly,
the ferromagnetic phase appears for a much larger range of parameters in
the replica symmetric solution than in the 1RSB solution. 

\subsubsection{3-regular graphs are 5-circular colorable, and require full step replica symmetry breaking}

We find that a typical random $3$-regular graph is $5$-circular colorable and hence provide evidence for Ne\v{s}et\v{r}il's Pentagon Conjecture. The corresponding 1RSB solution is interesting because it does not feature any frozen variables and is unstable towards further steps of RSB (we will give precise definitions of these terms in the later parts of the manuscript).

Among discrete models with known ground state energy (zero energy in this case) this is the only case we know of where the 1RSB solution is not stable. Experience with other (mostly dense) systems suggests that in such a case the full-step replica symmetry breaking solution is probably the correct one. Of course many problems, including the random satisfiability or coloring deep in the unsatisfiable regime are known to require the full-RSB (FRSB) approach, but in all these known examples the determination of the ground state energy itself is non-trivial. Knowing the ground state energy might considerably simplify the theoretical development. 

The RSB hierarchy inspired a lot of mathematical research and a number of very remarkable results were obtained for cases where the 1RSB picture is correct and the corresponding Gibbs states can be described in terms of frozen variables. The most remarkable being perhaps the proof of the $K$-satisfiability threshold at large $K$ \cite{ding2014proof}. Current mathematical tools are much weaker for the cases without frozen variables. And for cases of diluted systems where 1RSB is not the correct description even the heuristic tools of physics were not yet developed into a useful form. In this sense the case of circular 5-coloring of 3 regular random graphs provides an important example where the very meaning of full-RSB in the diluted systems can be investigated, both heuristically and rigorously. This is perhaps the most important result of the present paper. 

Remarkably although the statistical description of the space of proper coloring seems challengingly hard for the case of 5-circular coloring of 3-regular random graphs, the problem is algorithmically easy. We show empirically that simple simulated annealing finds proper coloring assignments even for very large systems. This suggests that a constructive (algorithmic) proof of 5-circular colorability of 3-regular graphs might be much easier than a non-constructive probabilistic proof. Further numerical investigation of this system might also shed light on the very nature of FRSB in diluted systems (again taking advantage of the fact that in the present case we know the exact ground state energy). 

\subsubsection{Combinatorially involved structure of survey propagation}

Another property that appears in circular coloring and none of the previous works we know of is a rather complex structure of the warning propagation equations that stand at the basis of survey propagation, a procedure used to determine the satisfiability and colorability threshold in \cite{MezardParisi02,MuletPagnani02}. In all the models where survey propagation was derived and applied, the variables were binary in which case there are only three options in terms of frozen variables: either the variable is frozen into value 1, or value 0 or none of them. Cases where variables are from a larger domain were not much studied. With the exception of the canonical coloring where the system of warnings is very simple and only $q$ different warning need to be considered. In circular coloring (and more generic problems with multi-valued variables) the structure of the warnings is more complex and the survey propagation equations are not very explicit. This difference is rooted in the nature of the constraints  that define the circular coloring problem and their investigation is another interesting result of the present work.

%%%%%%%%%%%%%%%% METHODS %%%%%%%%%%%%%%%%

\section{Cavity Method for Circular Coloring}

We apply the cavity method as developed in \cite{MezardParisi01,MezardParisi03}. This section mostly presents the  general formalism. The next section summarizes the results. 

We define circular coloring in terms of its Hamiltonian which can be expressed as
\begin{equation}
 \mathcal{H}(\mathbf{s})  =  \sum_{(i,j)\in\mathcal{E}} H(s_i,s_j) = \sum_{(i,j)\in\mathcal{E}}
 ( 1 - \delta_{s_i, s_j -1} - \delta_{s_i, s_j +1} )\, ,  \label{eq: H(si,sj)}
\end{equation}
where the algebra in the indices is modulo $q$.

\subsection{Replica symmetric solution}

The Hamiltonian (\ref{eq: H(si,sj)}) does only contain pairwise interactions and for generic pairwise models the belief propagation (BP) equations read
\begin{equation}
\psi_{s_{i}}^{i\rightarrow j} =
\frac{1}{Z^{i\rightarrow j}}
\prod_{k\in N(i) \backslash j} \sum_{s_k} \Phi_{i,k}(s_i , s_k) \ \psi_{s_{k}}^{k\rightarrow i} \, .
\label{eq:generic_BP_rekursion}
\end{equation}
Here $\psi_{s_{i}}^{i\rightarrow j}$ is the marginal probability for node $i$ to be of color $s_i$ if node $j$ is absent and $\Phi_{i,k}(s_i , s_k)$ is the interaction between pairs of nodes $i$ and $k$ and $Z^{i\rightarrow j}$ is guaranteeing the normalization. For a detailed derivation of a generic form see \cite{MezardMontanari07}. 

Given the Hamiltonian (\ref{eq: H(si,sj)}) and the corresponding Gibbs measure $\mu(\mathbf{s})=\frac{1}{Z}e^{-\beta\mathcal{H}(\mathbf{s})}$ the interaction term becomes $\Phi_{i,k}(s_i , s_k) = \frac{1}{Z}e^{-\beta H(s_i,s_k)}$ and the BP equations can be simplified. For $q>2$ they become
\begin{equation}
\psi_{s_{i}}^{i\rightarrow j} \equiv
\mathcal{F}(\{\mathbf{\psi}^{k\rightarrow i}\})=
\frac{1}{Z^{i\rightarrow j}}\prod_{k\in N(i)\backslash
  j}[e^{-\beta}+(1-e^{-\beta})(\psi_{s_{i}+1}^{k\rightarrow
  i}+\psi_{s_{i}-1}^{k\rightarrow i})]\, .\label{eq:BP rekursion}
\end{equation}
The term multiplied by $1-e^{-\beta}$ assigns more probability when neighbors have indeed consecutive colors.  The fixed point of the above equations can be used to compute the corresponding marginal distributions
\begin{equation}
\psi_{s_{i}}^{i} =
\frac{1}{Z^{i}}\prod_{k\in N(i)}[e^{-\beta}+(1-e^{-\beta})(\psi_{s_{i}+1}^{k\rightarrow
  i}+\psi_{s_{i}-1}^{k\rightarrow i})]\, ,\label{eq:BP_marginal}
\end{equation}
An important property of the above equations is that they provide exact marginals on trees and hence serve as a basis for an asymptotically exact solution on locally tree-like graphs, such as random graphs of fixed average degree. The BP messages do then allow access to the thermodynamic quantities. In the Bethe approximation the free energy $F$ (free entropy $-\beta F$) decomposes into its node and edge contribution $Z^{i}$ and $Z^{ij}$ respectively
\begin{eqnarray}
-\beta F & = & \sum_{i}\log Z^{i}-\sum_{(ij)\in \mathcal{E}} \log Z^{ij}\nonumber \\
 & = & \sum_{i}\log\left[\sum_{s=1}^{q}\prod_{k\in N(i)}[e^{-\beta}+(1-e^{-\beta})(\psi_{s+1}^{k\rightarrow i}+\psi_{s-1}^{k\rightarrow i})]\right]\nonumber \\
 &  & -\sum_{(ij)\in \mathcal{E}}\log\left[e^{-\beta}+(1-e^{-\beta})\sum_{s}\psi_{s}^{i\rightarrow j}(\psi_{s+1}^{j\rightarrow i}+\psi_{s-1}^{j\rightarrow i})\right]\,.\label{eq:bethe_free_entropy}
\end{eqnarray}
And from the Legendre transformation of the free energy we may then also derive other thermodynamic quantities, such as the energy $E$ and entropy $S$
\begin{equation}
-\beta F(\beta)=-\beta E+S(E)\,.\label{eq:F=E-TS}
\end{equation}
For instance, the energy is obtained as $E=\partial_{\beta}\left[\beta F\right]$. 

To examine the validity of the RS solution one can investigate when it loses its stability towards small perturbations in the messages. This is equivalent  to the non-convergence of BP on a single graph. The stability can be deduced from the leading eigenvalue of the Jacobian of the outgoing messages $\psi_{\tau}^{1\rightarrow0}$ when the incoming message $\psi_{\sigma}^{2\rightarrow 1}$ is varied infinitesimally.
\begin{equation}
J^{\tau\sigma}=\frac{\partial\psi_{\tau}^{1\rightarrow 0}}{\partial\psi_{\sigma}^{2\rightarrow 1}}\Big|^{\text{RS}}.\label{eq:Jacobian}
\end{equation}
For $q\ge 3$, the entries of the Jacobian in terms of the BP messages read
\begin{eqnarray}
J^{\tau \sigma} & = & (1-e^{-\beta}) \, \psi_{\tau}^{1\rightarrow 0} \left\{  \frac{\left( \delta_{\sigma,\tau-1} + \delta_{\sigma,\tau+1} \right)}{e^{- \beta} + (1-e^{- \beta})(\psi_{\tau-1}^{2\rightarrow 1}+\psi_{\tau+1}^{2\rightarrow 1})}  \right. \\
 &  & \left. - \frac{\psi_{\sigma + 1}^{1\rightarrow 0}}{e^{- \beta} + (1-e^{- \beta})(\psi_{\sigma}^{2\rightarrow 1}+\psi_{\sigma+2}^{2\rightarrow 1})} -  \frac{\psi_{\sigma - 1}^{1\rightarrow 0}}{e^{- \beta} + (1-e^{- \beta})(\psi_{\sigma-2}^{2\rightarrow 1}+\psi_{\sigma}^{2\rightarrow 1})}  \right\}
\, .
 \label{eq:J_general}
\end{eqnarray} 
The explicit eigenvalues will be given in the results section.

\subsection{One Step Replica Symmetry Breaking \label{sub:1RSB-framework}}

The correctness of the cavity method is conditioned on the fact that two-point correlations decay fast. In many diluted systems this condition does not hold and it is required to extend the replica symmetric version of the cavity method to compute the \emph{correct} thermodynamic properties of the system. Formally, such an extension is necessary when the Gibbs measure loses its extremality \cite{MezardMontanari07,ZdeborovaKrzakala07}. The extremality can sometimes be restored by decomposing the measure into a set of pure states which we will also refer to as "clusters". In a later section we will compute the spin glass temperature from the eigenvalues of (\ref{eq:Jacobian}) as the point below which the RS assumption is not valid. To take the clustered structure of the phase space into account, the replica symmetry breaking version of the cavity method was developed in \cite{MezardParisi01,MezardParisi03}. In the framework of \emph{one step} replica symmetry breaking (1RSB) each state corresponds to a different fixed point of the BP equations. The 1RSB  equations then deal with weighted averages over all such states and read 
\begin{eqnarray}
P^{i\rightarrow j}(\mathbf{\psi}^{i\rightarrow j}) & = &
\frac{1}{{\cal Z}^{i\rightarrow j}}\int\prod_{k\in N(i)\backslash j}\textrm{d}\mathbf{\psi}^{k\rightarrow i}P^{k\rightarrow i}(\mathbf{\psi}^{k\rightarrow i})\delta(\mathbf{\psi}^{i\rightarrow j}-\mathcal{F}(\{\mathbf{\psi}^{k\rightarrow i}\}))\,\left(Z^{i\rightarrow j}\right)^{m}\,,\label{eq:1RSB recursion}
\end{eqnarray}
where the function ${\cal F}$ and the term $Z^{i\rightarrow j}$ are
defined by the replica symmetric equation (\ref{eq:BP rekursion}), and ${\cal
  Z}^{i\rightarrow j}$ is a normalization constant. 

Likewise the thermodynamic quantities must be re-evaluated. For that purpose the \emph{replicated free energy} $\Phi(\beta,m)$ is introduced as follows 
\begin{equation}
e^{-\beta mN\Phi(\beta,m)}=\sum_{\{\mathbf{\psi}\}}Z(\{\mathbf{\psi}\})^{m}=\sum_{\{\mathbf{\psi}\}}e^{-\beta mNf(\{\mathbf{\psi}\})}=\int\textrm{d}fe^{-N[\beta mf(\beta)-\Sigma(f)]},\label{eq:Z(beta,m)}
\end{equation}
where the sum is over all different BP fixed points ${\{\mathbf{\psi}\}}$, each corresponding to a cluster, $\Sigma(f)$ counts the number of clusters having Bethe free energy $f$ and is usually referred to as complexity. The previous equation can be evaluated at its saddle point in the thermodynamic limit $N\rightarrow\infty$ and we obtain
\begin{equation}
-\beta m\Phi(\beta,m)=-\beta mf+\Sigma(f).\label{eq:Legendre replicated free entropy}
\end{equation}
The replicated free energy $\Phi$ is the Legendre transformation of $\Sigma$ and hence the following identities hold
\begin{equation}
f=\partial_{m}\left[m\Phi(\beta,m)\right],\hspace{0.5cm}\Sigma=\beta m^{2}\partial_{m}\Phi(\beta,m),\hspace{0.5cm}\beta m=\partial_{f}\Sigma(f)\,.\label{eq:Legendre identities}
\end{equation}
The Bethe approximation can then be adopted to compute the 1RSB free energy $\Phi = \sum_{i\in\mathcal{V}}\Delta\Phi^{i}-\sum_{(ij)\in\mathcal{E}}\Delta\Phi^{ij}$ for which the node and edge terms read
\begin{equation}
\Delta\Phi^{i,\,ij}=-\frac{1}{\beta m}\ln\int\prod_{a=1}^{l}\textrm{d}\mathbf{\psi}^{a}P(\mathbf{\psi}_{a})\,\delta\left(\mathbf{\psi}-\mathcal{F}(\{\mathbf{\psi}^{a}\})\right)\,\left(Z^{i,\,ij}\right)^{m}\,,\label{eq:replicated free energy}
\end{equation}
where $l$ is the number of neighbors on which $Z^i$ and $Z^{ij}$ depend. The 1RSB equations can be solved efficiently using population dynamics as introduced in \cite{MezardParisi01}: $P^{i\rightarrow j}(\mathbf{\psi}^{i\rightarrow j}) $ is approximated by a population of messages that is updated according to (\ref{eq:1RSB recursion}). After convergence, the population $\{\psi^*\}$ is an approximation of the true set of fixed points $\{\psi\}$.

Following the same line of arguments as for the replica symmetric
solution, we investigate the stability of the 1RSB solution towards
further steps of replica symmetry breaking. Small
perturbations can occur in the 1RSB solution either in $P(\mathbf{\psi})$ \emph{or} in
$\mathbf{\psi}$. In this work only the second case is investigated and
will already suffice to exclude the exactness of 1RSB. A description
of this stability analysis was for instance given in \cite{zdeborova2009statistical}. Numerically, one first waits $\tau$ iterations until the 1RSB equations converged for a given value of the re-weighting parameter $m$ towards $\{\mathbf{\psi^{*}}\}$. Then the population is duplicated and a small noise is introduced $\{\mathbf{\psi}^{*}+\delta\mathbf{\psi}^{*}\}$. Subsequently $t$ further iterations are performed. If both populations do not converge towards the same fixed point in the limit of many iterations, the 1RSB solution is said to be unstable. For a given $m$ the convergence can be tracked by means of the evolution of the noise $Q_{m}(t) = \sum_{\{\mathbf{\psi}^{*}\}}\left|\delta\mathbf{\psi}_{\tau+t}^{*}\right|.$ And if $\lim_{t\rightarrow\infty}Q_{m}(t)\not\to0$ the solution is  unstable towards further steps of RSB. The value of $m$ at which the instability sets in will be denoted by $m^{\Delta}$.

\section{Cavity method at zero temperature \label{sec:cavity_zero_temperature}}

In circular coloring the ground state quantities are particularly interesting and therefore the zero temperature limit of the 1RSB equations should be considered. There are complementary ways to take the zero temperature limit that we shall now briefly recall. 

If one is merely interested in proper assignments for which the total
energy is zero one considers the so-called {\it entropic zero-temperature limit} \cite{mezard2005landscape,zdeborova2009statistical}. If $e=0$ then the free energy reads $-\beta f=s$ and can be computed via (\ref{eq:bethe_free_entropy}). With this, and after defining $\Phi_{\mathrm{s}}(m)\equiv-\beta m\Phi(\beta,m)\vert_{\beta\rightarrow\infty}$, the replicated free entropy from (\ref{eq:Legendre replicated free entropy}) might be rewritten as $\Phi_{s}(m)=ms+\Sigma(s)$.  The replicated entropy $\Phi_{s}(m)$ can then be computed from (\ref{eq:replicated free energy}) and hence the number of clusters of size $s$ of proper assignments $\Sigma(s)$ is obtained. With the Legendre properties, the remaining free parameter $m$ can be varied in order to access $\Sigma(s)$ over the whole range. 

On the other hand, in the energetic zero temperature limit we take $\beta\rightarrow\infty$ while $y\equiv m\beta$ remains finite. In this case the free energy equals the energy, i.e. $f=e$ and from (\ref{eq:Legendre replicated free entropy}) one obtains $-y\Phi(y)=-y\,e+\Sigma(e).$ It's hence called \emph{energetic zero temperature limit} and allows us to compute the ground state energy. In this limit the structure of the messages alter considerably as developed in the work of \cite{MezardParisi03}.

\subsection{Warning Propagation \label{sub:Warning_Propagation}}

To derive the energetic zero temperature 1RSB analysis, one starts
with the \emph{warning propagation}, which is a zero
temperature limit of the belief propagation
(\ref{eq:BP rekursion}). One introduces the cavity fields $h_{s_i}^{i\rightarrow j}$ as $\psi_{s_i}^{i\rightarrow j} \equiv \exp(-\beta h_{s_i}^{i\rightarrow j})$ and the sum in the generic BP equations
(\ref{eq:BP rekursion}) is replaced by taking the maximal marginal
i.e. $\sum_{s_k} \rightarrow \max_{s_k}$. 
Adapting the generic BP equations (\ref{eq:generic_BP_rekursion}) accordingly yields $\psi_{s_i}^{i\rightarrow j} \equiv \exp(-\beta h_{s_i}^{i\rightarrow j}) \cong \exp (-\beta \sum_{k\in \partial i\backslash j} \min_{s_k} [\mathcal{H}(s_i,s_k)+h_{s_k}^{k\rightarrow i}])$ and therefore, after taking the logarithm the equations read
\begin{equation}
h_{s_i}^{i\rightarrow j}=\sum_{k\in \partial i \backslash j} \min_{s_k} \left[ \mathcal{H}(s_i,s_k)+h_{s_k}^{k\rightarrow i} \right]-\min_{s_k} h^{k \rightarrow i}_{s_k}  \, ,
\label{eq:warning_propagation}
\end{equation}
The term on the right hand side can be interpreted as a \emph{warning} from node $k$, incoming to node $i$ -- let's denote it by $u_{s_k}^{k\rightarrow i}$. The sum over the warnings $\sum_{k\in \partial i \backslash j}u_{s_k}^{k\rightarrow i}$ yields the \emph{cavity field} acting on node $i$, assuming that node $j$ is absent and the above equation can be re-expressed as
\begin{equation}
 h_{\tau}^{i \rightarrow j}=\sum_{k \in N(i)\backslash j} u_{\tau}^{k\rightarrow i} \, ,
 \label{eq:h(u)}
\end{equation}
where the $\tau^{\tiny th}$ component of $u(h_1,\dots,h_\tau)$ reads 
\begin{equation}
\hat{u}_{\tau}(h_1,\dots,h_q)=\min\left(h_{1}+1,\dots,h_{\tau-2}+1,h_{\tau-1},h_{\tau}+1,h_{\tau+1},h_{\tau+2}+1,\dots,h_{q}+1\right)-\omega(h_1,\dots,h_q)\,.\label{eq:uhat(h)}
\end{equation}
with 
\begin{equation}
 \omega(h_1,\dots,h_q)=\min\left(h_{1},h_{2},\dots,h_{q-1},h_{q}\right)\label{eq:omega(h)}
\end{equation}
assuring that $u^{i\rightarrow j}_{s_k} = \{0,1\}$. We identify a contradiction or \emph{energy shift} along the directed edge $i\rightarrow j$ as an assignment of $s_i$ such that $\mathcal{H}(s_i,s_j) = 1$. Then a contradiction is related to $u^{i\rightarrow j}_{s_i} = 1$. Contrarily, if $u^{i\rightarrow j}_{s_i} = 0$ no contradiction is caused by the assignment. The $\tau^{\tiny th}$ component of the field $h^{i\rightarrow j}$ is therefore related to the number of contradictions along all incoming edges caused by assigning color $\tau$ to node $i$ if the edge $j$ is absent. Accordingly, an \emph{energy shift} can be assigned to a single node: this is just the number of contradictions from \emph{all} incoming edges. Subsequently $q$-component vectors will be denoted by bold symbols $(h_1,h_2,\dots,h_q)\equiv \mathbf{h}$ to simplify the notation.

\subsection{Survey Propagation \label{sub:Energetic-Zero-Temperature-Limit}}

The reasoning of \cite{braunstein2003polynomial} was generic enough to be
applied to circular coloring as well (up to the point where the list
of relevant warnings is explicated). This is because all the
information on the Hamiltonian is absorbed in
$\omega(\mathbf{h})$. Therefore the 1RSB equation (\ref{eq:1RSB
  recursion}) reads (up to a normalization, recalling $y=m\beta$)
\begin{eqnarray}
 P_{i\rightarrow j}(\mathbf{h})&\cong&\int\prod_{k\in N(i)\backslash
   j}\mathrm{d}\mathbf{u}_{k}Q_{k\rightarrow
   i}(\mathbf{u}_{k})\,\delta\left(\mathbf{h}-\sum_{k\in
     N(i)\backslash
     j}\mathbf{u}_{k}\right)\,\exp\left[-y\,\omega\left(\sum_{k\in
       N(i)\backslash j}\mathbf{u}_{k}\right)\right],\label{eq:P(h)} \\
 Q_{i\rightarrow j}(\mathbf{u})&=&\int\mathrm{d}\mathbf{h}P_{i\rightarrow j}(\mathbf{h})\,\delta\left(\mathbf{u}-\hat{\mathbf{u}}(\mathbf{h})\right)\,.
 \label{eq:Q(u)}
\end{eqnarray}
The above equations are referred to as the SP-y solution, or as the
energetic zero temperature limit of the cavity solution. 

At this point another limit can be taken, namely $y\rightarrow\infty$, which prohibits any kind of contradiction as only those local fields $\mathbf{h}=\sum_{k}\mathbf{u}_{k}$ contribute in (\ref{eq:P(h)}) that contain at least one zero component, i.e. no contradiction. This is due to the re-weighting term $\exp\left[-y\omega(\mathbf{h})\right]$ that, in this limit, only contributes when the energy is zero. The resulting equations are known as \emph{survey propagation}. In this limit we can characterize the distributions over the warnings $Q(\mathbf{u})$ by the set of parameters $\{\eta\}$ that are associated with the different possible warnings $\{\mathbf{u}\}$. For each distinct warning $\mathbf{u}^{(i)}$ we have $\eta^{(i)} = Q(\mathbf{u}^{(i)})$. In other words, solving the equations (\ref{eq:P(h)}) and (\ref{eq:Q(u)}) amounts to finding the equivalence classes $[\mathbf{h}]_{i}$ that map onto the warning $\mathbf{u}^{(i)}$, i.e. $[\mathbf{h}]_{i}=\{\mathbf{h}\in\{\mathbf{h}\}\,|\,\mathbf{\hat{u}}(\mathbf{h})=\mathbf{u}^{(i)}\}$. This can be written in a recursion for the parameters $\eta^{(i)}$, where $(i)$ indicates the $i^{\tiny th}$ equivalence class and $i$ runs over all integers from $0$ to $\left|\{\mathbf{u}\}\right|-1$
\begin{equation}
\eta_{i\rightarrow j}^{(\alpha)}\cong\sum_{\pi(\{\alpha_{k}\})}\mathbb{I}\left(\sum_{k\in N(i)\backslash j}\mathbf{u}_{k}^{(\alpha_{k})}\,\in[\mathbf{h}]_{\alpha}\right)\prod_{k\in N(i)\backslash j}\eta_{k\rightarrow i}^{(\alpha_{k})}\,.\label{eq:closed form spy}
\end{equation}

In contrast to previously studied cases of the survey propagation
equations, e.g. \cite{BraunsteinMezard05,MuletPagnani02},
(\ref{eq:closed form spy}) has a rather complicated structure due to
the diverse cases that are to be distinguished. Unlike regular
coloring, circular coloring is much stronger constrained: If one fixes
the color of one node to $s$, all nearest neighbors must take either
of the two colors $s-1$ or $s+1$, second-nearest neighbors are restricted to $s$, $s-2$ and $s+2$ and so forth. In fact up to $k(q)$-nearest neighbors are restricted. Exemplary: for $q=5$ we have $k=3$.
Consequently the closure of (\ref{eq:closed form spy})
is much more involved and no explicit formula was obtained here for
the generic case. Studying the structure of the warnings $\mathbf{u}$
and fields $\mathbf{h}$ for the $5$-circular coloring on $3$-regular graphs in the section~\ref{sub:5cc} illustrates the difference further. But even if the closed form is unknown, the equations can either be generated numerically in an exhaustive approach, or they can simply be solved by computing (\ref{eq:P(h)}) and (\ref{eq:Q(u)}) in the limit $y\to\infty$ with the population dynamics.

When aiming to close the equations on $\mathbf{u}$, we must find the
mapping of incoming fields to the outgoing warnings and identify all
fields that cause the same response in (\ref{eq:uhat(h)}). Without loss
of generality we can identify $\min(\mathbf{h})$ with
$h_{\arg\min(\mathbf{h})}=0$ and all other components
$i\neq\arg\min(\mathbf{h})$ with $h_i=1$ because they cause the same
respond in (\ref{eq:uhat(h)}). After this identification we must still 
in general distinguish $\left|\{\mathbf{h}\}\right|=2^q$ fields.

Let's assume that the rotational and reflectional symmetry of
$q$-circular coloring is not broken below a certain point $q_\text{sym}(d)$. For $q<q_{\text{sym}}(d)$ we can take advantage of this symmetry and $\left|\{\mathbf{h}\}\right|<2^q$ cases must be distinguished for the incoming fields -- when we count the different fields and warnings it is understood that e.g. $(h_1,0,0,0,0)$ and $(0,0,h_3,0,0)$, with $h_1$ not necessarily equal to $h_3$, is counted only once. The distinguished cases $\left|\{\mathbf{h}\}\right|$ are then obtained by counting in how many different ways $q$ beads can be circularly connected when each bead might be of either of two colors and rotations and reflections of this circular string are regarded as equivalent. These objects are known as \emph{bracelets} in combinatorics and they are counted by
\[
|\{\mathbf{h}\}|=\frac{1}{2q}\sum_{t|q}\phi(t)\,2^{\frac{q}{t}}+\begin{cases}
2^{\frac{q-1}{2}} & q\ \textrm{odd}\\
3\cdot2^{\frac{q}{2}-2} & q\ \textrm{even}
\end{cases}\,.
\label{eq:bracelets}
\]
where the sum goes over all divisors of $q$ and $\phi(t)$ is Euler's
Totient function that counts how many integers are smaller or equal to $t$
and share no common positive divisors with $t$ except 1. Since (\ref{eq:uhat(h)}) introduces a non-injective surjection on $\{\mathbf{h}\}$, it is more involved to count the cardinality of $\{\mathbf{u}\}$ under consideration of the symmetry. However, it is simple to provide the first few values of the sequence from numerical evaluation, done in table \ref{tab:cardinality u and h}.

\begin{table}
\begin{tabular}{|c|c|c|}
\hline 
$q$ & $|\{\mathbf{h}\}|$ & $|\{\mathbf{u}\}|$\tabularnewline
\hline 
\hline 
$3$ & $3$ & $2$\tabularnewline
\hline 
$4$ & $5$ & $2$\tabularnewline
\hline 
$5$ & $7$ & $4$\tabularnewline
\hline 
$6$ & $12$ & $6$\tabularnewline
\hline 
$7$ & $17$ & $8$\tabularnewline
\hline 
$8$ & $29$ & $13$\tabularnewline
\hline 
$9$ & $45$ & $17$\tabularnewline
\hline 
$10$ & $77$ & $26$\tabularnewline
\hline 
$11$ & $125$ & $36$\tabularnewline
\hline 
$12$ & $223$ & $56$\tabularnewline
\hline 
$13$ & $379$ & $82$\tabularnewline
\hline 
$14$ & $686$ & $128$\tabularnewline
\hline 
\end{tabular}\protect\caption{The cardinality of the set of all possible fields and warnings in the paramagnetic phase where vectors that are equivalent under the group actions of rotation and reflection are only counted once. \label{tab:cardinality u and h}}
\end{table}

Note that this is intrinsically different from e.g. regular coloring where, after taking into account the permutation symmetry, only two generic cases must be distinguished: the case in which $\mathbf{h}$ possesses a unique minimum, causing $\mathbf{u}=\mathbf{e}_{\tau}$, where $\mathbf{e}_{\tau}$ is a unit vector in direction $\tau$, and the case in which the minimum is degenerate, causing $\mathbf{u}=\mathbf{0}$. Such a simplification is not possible for circular coloring and per se $\left|\{\mathbf{u}\}\right|$ generic cases must be distinguished (following this section we will give a concrete example to illustrate the difference). Within the framework of circular coloring it is therefore sensible to slightly modify the notation of what is known as \emph{frozen} variables in other constraint satisfaction problems (CSPs), such as generic coloring or K-SAT. In previously considered CSPs the variables could either be trivial or point into one of the $q$ possible directions $\mathbf{e}_{\tau}$. As we have just seen in this section that is no longer the case in circular coloring. Instead of frozen variables the term \emph{confined variables} is therefore introduced in order to emphasis this different nature.

\subsection{Example of Survey Propagation for Five Circular Coloring \label{sub:5cc}}

In this section we consider the survey propagation equations of the
example of $5$-circular coloring of $3$-regular graphs, in this case
we can obtain explicit survey propagation equations. For this
example we shall confirm in section
\ref{sub:Energetic-Zero-Temperature-Limit-Results} that the symmetry
under rotation and reflection is not broken. 

\paragraph{First we derive the set of warnings.}
Assume that no neighbor imposes any constraint on node $i$,
i.e. $\mathbf{h}=(0,0,0,0,0)$, then the outgoing warning yields
$\mathbf{u}^{(0)}=(0,0,0,0,0)$. Similarly, if the neighbors of $i$ constrain
one color only, say $s_{i}=1$ then $\mathbf{h}=(h_{1},0,0,0,0)$ (with
$h_1>0$) yielding $\mathbf{u}^{(0)}$ again. The last case that yields
this ``zero'' warning would be
$\mathbf{h}=(h_{1},h_{2},0,0,0)$. Proceeding in the same fashion, one
might check all possible fields $\{\mathbf{h}\}$ -- the according
mapping from $\mathbf{h}$ to $\mathbf{u}$ can, again up to the
intrinsic symmetry, be found in table \ref{tab:h to u}.

\begin{table} 
		\begin{tabular}{ccc} 
			$\mathbf{h}$ &  & $\mathbf{u}$ 
			\tabularnewline
			 &  & 
			\tabularnewline \cline{1-1}  
			\multicolumn{1}{|c|}{$\begin{array}{c} (0,0,0,0,0)\\ (h_{1},0,0,0,0)\\ (h_{1},h_{2},0,0,0) \end{array}$} & $\rightarrow$ & $\mathbf{u}^{(0)} =(0,0,0,0,0)$
			\tabularnewline \cline{1-1}   
			&  & 
			\tabularnewline \cline{1-1}  
			\multicolumn{1}{|c|}{$\begin{array}{c} (h_{1},h_{2},h_{3},0,0)\\ (h_{1},0,h_{3},0,0) \end{array}$} & $\rightarrow$ & $\mathbf{u}^{(1)} =(0,1,0,0,0)$
			\tabularnewline \cline{1-1}   
			&  & 
			\tabularnewline \cline{1-1}  
			\multicolumn{1}{|c|}{$(h_{1},h_{2},h_{3},h_{4},0)$}
                        & $\rightarrow$ & $\mathbf{u}^{(2)}  = (0,1,1,0,1)$
			\tabularnewline \cline{1-1}   
			&  & 
			\tabularnewline \cline{1-1}  
			\multicolumn{1}{|c|}{$(h_{1},h_{2},0,h_{4},0$)}
                        & $\rightarrow$ & $\mathbf{u}^{(3)}  = (0,0,1,0,1)$
			\tabularnewline \cline{1-1}  
		\end{tabular} \par 
	\caption{All possible non-contradictory incoming fields $\mathbf{h}$, i.e. all those fields that contain at least one zero component, and their corresponding outgoing warnings $\mathbf{u}$ (up to relevant symmetry) \label{tab:h to u}} .
\end{table}  

\paragraph{Conversely the warnings map to the fields in a more tedious manner.}

Assuming $d=3$ and labeling the two incoming edges as 'left' and 'right'. One might obtain $\mathbf{u}^{(0)}$ by combining the two incoming warnings $\mathbf{u}^{(0)}$ and $\mathbf{u}^{(0)}$. Note that there is only one such combination because of the symmetry under exchanging left $\rightleftharpoons$ right. Another way to obtain $\mathbf{u}^{(0)}$ is by combining $\mathbf{u}^{(0)}$ from the left with $\mathbf{u}^{(1)}$ from the right and vice versa (no symmetry w.r.t left $\rightleftharpoons$ right). The last possible combination would then be $\mathbf{u}^{(1)}$, incoming from one of the two edges and $\mathbf{u}^{(1)}$ from the remaining edge, such that either $(h_{1},0,0,0,0)$ or $(h_{1},h_{2},0,0,0)$ is obtained, i.e. there are $3$ such combinations. For the last three cases we must also take into account the $5$ possible rotations. We summarize in the following table
%\vspace{5bp}
\begin{center}
\begin{tabular}{cccc}
left & right & left $\rightleftharpoons$ right & \tabularnewline
\hline 
$u^{(0)}=(0,0,0,0,0)$ & $u^{(0)}=(0,0,0,0,0)$ & yes & $\eta^{(0)}\eta^{(0)}$\tabularnewline
$u^{(0)}=(0,0,0,0,0)$ & $u^{(1)}=(1,0,0,0,0)$ + rot & no & $5\cdot 2 \eta^{(0)}\eta^{(1)}$\tabularnewline
$u^{(1)}=(1,0,0,0,0)$ & $u^{(1)}=(1,0,0,0,0)$ + rot & yes & $5\cdot \eta^{(1)}\eta^{(1)}$\tabularnewline
$u^{(1)}=(1,0,0,0,0)$ & $u^{(1)}=(0,1,0,0,0)$ + rot & no & $5\cdot 2 \eta^{(1)}\eta^{(1)}$\tabularnewline
\end{tabular}
\end{center}
\vspace{5bp}
Providing another example, namely the one on $\mathbf{u}^{(2)}$, one has the following possible combinations  
\vspace{5bp}
\begin{center}
\begin{tabular}{cccc}
left & right & left $\rightleftharpoons$ right & \tabularnewline
\hline 
$u^{(2)}=(1,1,0,1,0)$ & $u^{(1)}=(0,0,1,0,0)$ & no & $2\eta^{(1)}\eta^{(2)}$\tabularnewline
$u^{(2)}=(1,0,1,1,0)$ & $u^{(1)}=(0,1,0,0,0)$ & no & $2\eta^{(1)}\eta^{(2)}$\tabularnewline
$u^{(2)}=(1,0,1,1,0)$ & $u^{(2)}=(1,1,0,1,0)$ & no & $2\eta^{(2)}\eta^{(2)}$\tabularnewline
$u^{(2)}=(1,0,1,1,0)$ & $u^{(3)}=(0,1,0,1,0)$ & no & $2\eta^{(2)}\eta^{(3)}$\tabularnewline
$u^{(2)}=(1,1,0,1,0)$ & $u^{(3)}=(1,0,1,0,0)$ & no & $2\eta^{(2)}\eta^{(3)}$\tabularnewline
$u^{(3)}=(1,0,1,0,0)$ & $u^{(3)}=(0,1,0,1,0)$ & no & $2\eta^{(3)}\eta^{(3)}$\tabularnewline
\end{tabular}
\end{center}
\vspace{5bp}
Proceeding in the same fashion for $\mathbf{u}^{(1)}$ and
$\mathbf{u}^{(3)}$ we end up with an explicit form of the survey
propagation equations  (\ref{eq:closed form spy})  for
$5$-circular coloring of $3$-regular random graphs
\begin{eqnarray} 
	\eta^{(0)} & \cong & \eta^{(0)}\eta^{(0)}+5\cdot 2\cdot\eta^{(0)}\eta^{(1)}+5\cdot3\cdot\eta^{(1)}\eta^{(1)}\nonumber \\ 
	\eta^{(1)} & \cong & 2\cdot\eta^{(0)}\eta^{(3)}+2\cdot\eta^{(1)}\eta^{(1)}+6\cdot\eta^{(1)}\eta^{(3)}+2\cdot\eta^{(3)}\eta^{(3)}\nonumber \\ 
	\eta^{(2)} & \cong & 4\cdot\eta^{(1)}\eta^{(2)}+2\cdot\eta^{(2)}\eta^{(2)}+4\cdot\eta^{(2)}\eta^{(3)}+2\cdot\eta^{(3)}\eta^{(3)}\nonumber \\ 
	\eta^{(3)} & \cong & 2\cdot\eta^{(0)}\eta^{(2)}+6\cdot\eta^{(1)}\eta^{(2)}+4\cdot\eta^{(1)}\eta^{(3)}+\eta^{(2)}\eta^{(2)}+4\cdot\eta^{(2)}\eta^{(3)}+2\cdot\eta^{(3)}\eta^{(3)}\,,
\label{eq:SP53} 
\end{eqnarray} 
where the right- and left hand side indicate iteration $k$ and $k+1$ respectively. Again, assuming the symmetry the normalization is given as $\eta^{(0)}+q\cdot\sum_i\eta^{(i)}$.

\section{Results for Regular Random Graphs}

We present the results for regular random graphs, i.e. every vertex
having the same degree. The neighborhood of almost every node in regular
random graphs looks the same \cite{Bollobas01} up to a distance $\Omega(\log{N})$, we
therefore assume that the cavity message is the same for every
node/edge. This simplifies considerably the numerical analysis of the
corresponding fixed point equations.

\subsection{Replica Symmetric Phase Diagram}

In this section we investigate the solution of the replica symmetric
BP equations on $d$-\textit{regular} random graphs. In this case the
message on every edge is equal and a physical solution of (\ref{eq:BP rekursion}) must be uniform over the edges. Accordingly we can simplify (\ref{eq:BP rekursion}) as
\begin{equation}
\psi_{s} = \mathcal{F}(\{\psi\}) =
\frac{1}{Z}[e^{-\beta}+(1-e^{-\beta})(\psi_{s+1}+\psi_{s-1})]^{d-1} \, . \label{eq:BP_regular}
\end{equation}
We associate the Bethe free energy (\ref{eq:bethe_free_entropy}) to each fixed point of (\ref{eq:BP_regular}) and the equilibrium solution is then obtained by selecting the fixed point that has minimal free energy.

\begin{figure}
\includegraphics[scale=0.5]{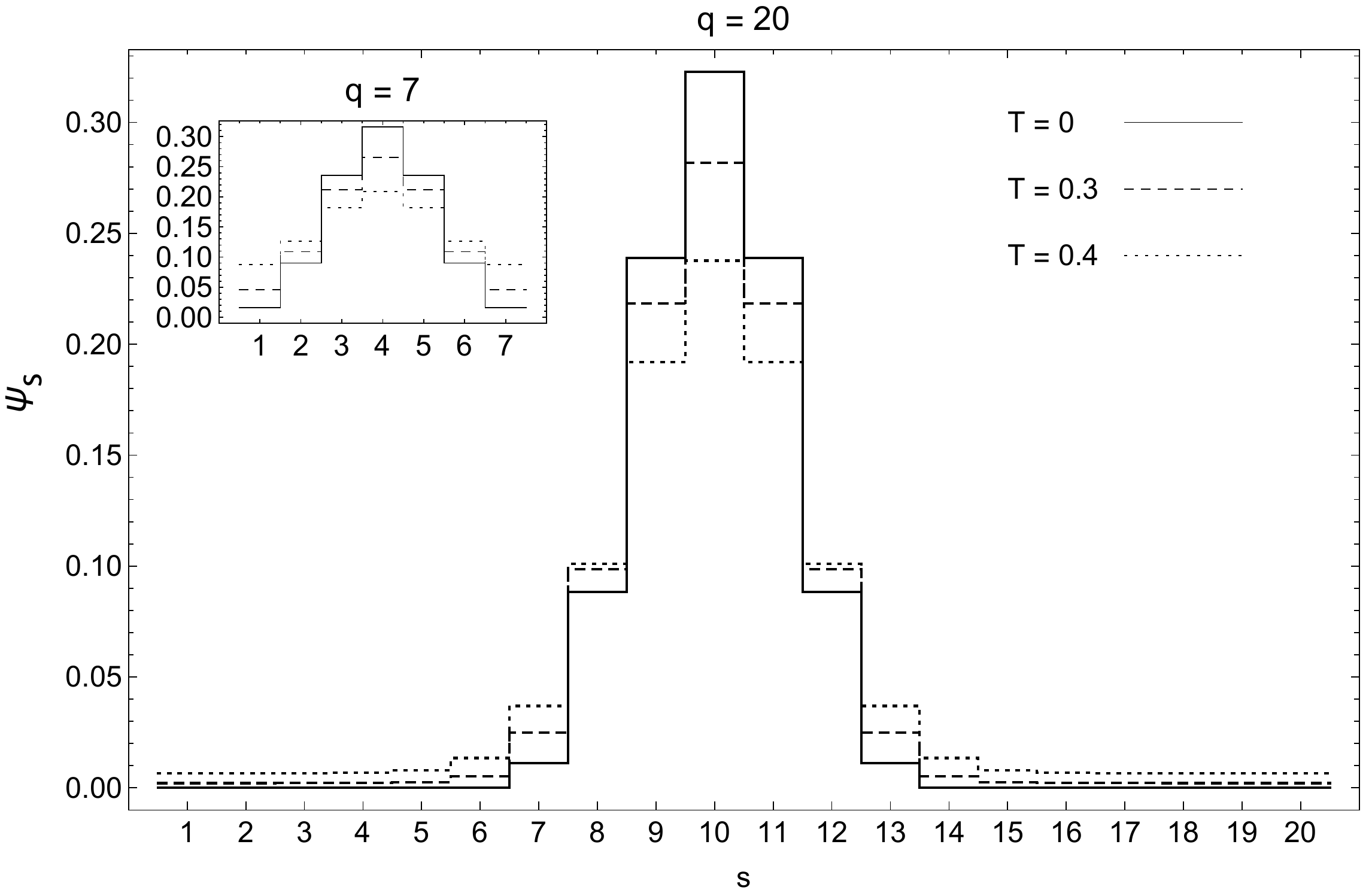}
\caption{Exemplary ferromagnetic replica symmetric solutions for $q=20$
  ($q=7$ in the inset) circular coloring on $3$-regular random
  graphs. We observed that roughly for $q\geq 11$
  the RS zero temperature solutions are quasi indistinguishable. 
\label{fig:ferro_sln} }
\end{figure}

One solution is the paramagnetic one, with a corresponding fixed point 
\begin{equation}
\psi_{s_{i}}^{i\rightarrow j} = \frac{1}{q}   \quad {\rm for \, \, all}
\quad s_i, (i,j)\in \mathcal{E} \, . 
\end{equation}
However, for sufficiently large values of $q$ we find another stable fixed
point that has ferromagnetic character. In this ferromagnetic fixed
point the probability to have a
given color concentrates in few adjacent colors. Examples of the form
on this ferromagnetic fixed point for random
$3$-regular graphs are depicted in Fig. \ref{fig:ferro_sln}. 
We verified by careful numerical search over initializations that no other fixed points
appear for the cases we studied.

In order to draw the replica symmetric phase diagram, we evaluate the
linear stability of the fixed points of (\ref{eq:BP_regular}) via the
spectrum of the linearization matrix $\cal T$

\begin{equation}
\left.{\cal T}^{\tau \sigma} = \frac{\partial\psi_{\tau}}{\partial\psi_{\sigma}}\right|^{\text{RS}} = (d-1)(1-e^{-\beta})\left\{  \chi_{\tau}  \left( \delta_{\tau-1,\sigma} +\delta_{\tau+1,\sigma} \right) - \left( \chi_{\sigma+1}+ \chi_{\sigma-1} \right) \psi_{\tau} \right\}
\label{eq:num_stability_jacobian}
\end{equation}
where $\chi_{\tau}=\frac{\Psi_{\tau}}{e^{-
    \beta}+(1-e^{-\beta})(\Psi_{\tau-1}+\Psi_{\tau+1})}$. Concerning
the instability of the paramagnetic solution towards the ferromagnet
we look for the smallest temperature $T_\mathrm{S}$ at which  the largest positive eigenvalue of the
matrix ${\cal T}$ is smaller than one. The negative eigenvalues of
${\cal T}$ indicate an anti-ferromagnetic instability which appears on
trees and other bipartite graphs, but that is unphysical on random
graphs, therefore the negative eigenvalues do not lead to a physical instability. 

For the paramagnetic fixed point with $\Psi_{\tau} = 1/q$ the
matrix ${\cal T}$ is a circulant matrix and the non-zero eigenvalues $\nu_j$ read
\begin{equation}
\nu_j =  (d-1) \frac{2}{q} \frac{1-e^{-\beta}}{e^{-\beta}+(1-e^{-\beta})\frac{2}{q}} \cos \frac{2\pi j}{q}  \ ,\   j=1,\dots,q-1 \, .
\label{eq:ev_numeric}
\end{equation}
The largest positive eigenvalue is $\nu_1$, the paramagnetic phase is
stable if and only if $\nu_1 < 1$. At zero temperature $\beta
\to \infty$ the degree of the graph needs to be larger than
$d>d_F$ where 
\begin{equation}
d^{\rm RS}_{\rm F}(q) = 1+1/\cos(2\pi /q)
\end{equation}
for the ferromagnetic instability to
appear. Analogously we will call $q^{\rm RS}_{\rm F}(d)$ the largest value
of $q$ for which the replica symmetric calculations predicts existence
of the ferromagnetic phase. When this condition is satisfied the paramagnetic solution
becomes unstable towards the ferromagnet for temperatures lower than
the so-called paramagnetic spinodal temperature 
\begin{equation}
T_{\mathrm{SP}} = 1 /  \ln \left[    {  1 + \frac {q}{ 2 \left( (d-1)\cos\frac{2\pi}{q}  -1 \right) }   }    \right] \, .
\label{eq:T_SP}
\end{equation}

\begin{figure}
\includegraphics[scale=0.5]{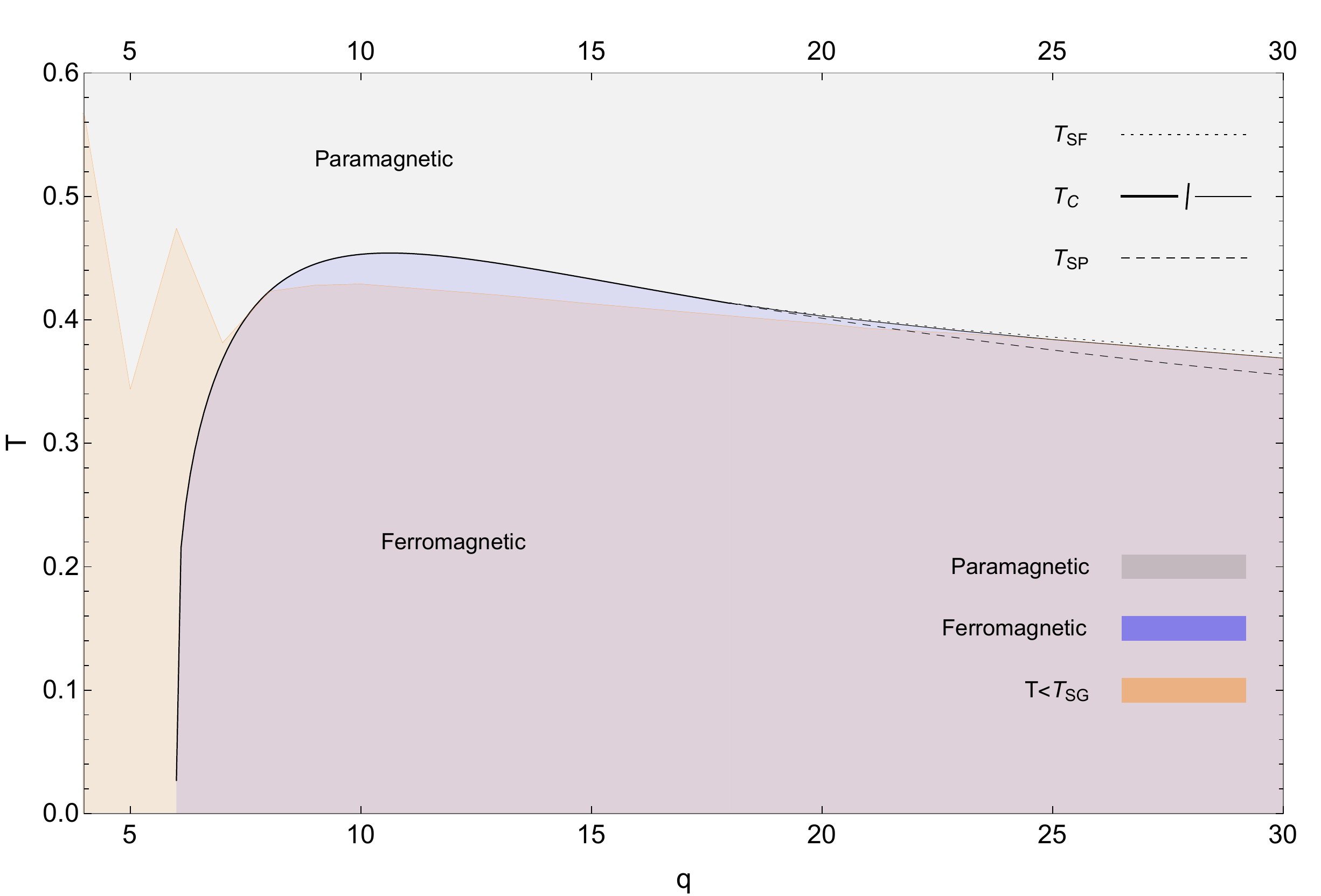}
\caption{The replica symmetric phase diagram for circular coloring of
  $3$-regular random graphs as a function of the number of colors $q$
  and the temperature $T$. We plotted the critical temperature
  $T_{\mathrm{c}}$, the spinodal lines $T_{\mathrm{SP}}$ and
  $T_{\mathrm{SF}}$ as the limits of stability of the para-
  and ferromagnetic solution respectively, and the spin glass
  stability which is the temperature below which the replica symmetric
  ansatz loses its validity. The critical temperature is plotted as
  thick (thin) solid line for $q<18$ ($q\geq18$). For $q<18$  the
  transition is second order and the spinodal line coincides with the
  critical temperature. For $q\geq 18$ the transition becomes first
  order and the two spinodal lines separate from the critical line and
  a regime of phase coexistence that grows with $q$ appears. 
\label{fig:rs_phase_diagram} }
\end{figure}

For random $3$-regular graphs we computed the detailed replica symmetric
phase diagram as a function of the number of colors $q$ and temperature $T$, depicted in Fig.~\ref{fig:rs_phase_diagram}. In
terms of the number of colors $q$ there are three different
regimes in the replica symmetric results. For $q<7$ the only solution for all temperatures is the
paramagnetic one. In the intermediate regime for $7\leq q \leq 18$ a
second order phase transition at $T_{\mathrm{c}}=T_{\mathrm{SP}}$ separates the ferromagnetic
phase (where the ferromagnetic solution is the only stable one)
from the paramagnetic phase (where the paramagnetic solution is the
only stable one). Eventually for $q>18$ the transition
becomes first order: the ferro- and paramagnetic phases coexist
for an interval of temperatures $T_{\mathrm{SP}} < T < T_{\mathrm{SF}}$. In this
regime of phase coexistence the thermodynamically correct solution is
the one with smallest free energy.  For $T<T_{\mathrm{c}}$ the ferromagnetic free energy is smaller than the paramagnetic one and hence favorable. At $T=T_{\mathrm{c}}$ the two free energies are crossing and for $T_{\mathrm{SF}}>T>T_{\mathrm{c}}$ both states still co-exist, but now the free energy of the paramagnetic state is smaller and therefore favorable.

The existence of the ferromagnetic phase can be
  understood intuitively. For a large number of colors it is very
  unlikely that two random colors will be adjacent, hence the problem
  is much more constrained than with fewer colors. The energetic gain
  from using only few colors that are close to each other on the cycle
  is overwhelming the entropic gain from using many colors for low
  enough temperature. 
In section \ref{sub:math} it was argued that circular coloring with
even $q$ can be reduced to $q=2$. Interestingly, the replica symmetric
investigation does not reproduce this fact.

\subsection{Spin-Glass Instability \label{sub:spinglass_stability}}

In order to investigate the spin-glass stability we must consider the Jacobian (\ref{eq:J_general}) and compute its leading eigenvalues. For the paramagnetic fixed-point the matrix is circulant and the computation can be done analytically. There are only two kind of entries:
\begin{equation}
J^{\tau\sigma}=\begin{cases}
(\frac{1}{q}-\frac{2}{q^{2}})\frac{1-e^{-\beta}}{e^{-\beta}+(1-e^{-\beta})\frac{2}{q}} & \text{if }\sigma=\tau\pm1\mod q\\
-\frac{2}{q^{2}}\frac{1-e^{-\beta}}{e^{-\beta}+(1-e^{-\beta})\frac{2}{q}} & \text{else}
\end{cases}
\end{equation}
and we obtain the following eigenvalues 
\begin{equation}
\lambda_{j}=\begin{cases}
0 & j=0\\
\frac{\cos\left(\frac{2\pi}{q}\,j\right)}{1+\frac{e^{-\beta}}{(1-e^{-\beta})}\,\frac{q}{2}} & j=1,2,\dots,q-1
\end{cases}\,.\label{eq:eigenvalues_jacobian}
\end{equation}
from which the leading eigenvalue is obtained by selecting 
\begin{equation}
j_{\text{max}} = \arg\max \left|\lambda_j\right|=\begin{cases}
q/2& q\ \textrm{even and }q\geq4\\
(q\pm1)/2 & q\ \textrm{odd and }q\geq 4
\end{cases} \, .
\label{eq:max. eigenvalue}
\end{equation}
With the stability condition $\kappa\lambda_{\rm max}^{2}<1$ \cite{Thouless86}, where $\kappa$ is the average excess degree, $\kappa=d-1$ for $d$-regular graphs, and $\kappa=c$ for Erd\H{o}s-R\'{e}nyi graphs, we obtain the spin glass temperature
\begin{equation}
T_{\mathrm{SG}}=
1/\ln\left(1+\frac{q}{2\left(\sqrt{d-1}|\cos\frac{2\pi}{q}j_{\text{max}}|-1\right)}\right)
\,.
\label{eq:T_c}
\end{equation}
For the ferromagnetic solution we rely on solving (\ref{eq:J_general})
numerically and then extracting $T_{\mathrm{SG}}$ as the point where
$(d-1) \lambda_{\rm max}^2 =1$, with $\lambda_{\rm max}$ being the
eigenvalue of the Jacobian that has largest absolute value. 
Note that the spin-glass transition can happen before, after or simultaneously with the ferromagnet transition.

\begin{figure}
\includegraphics[scale=0.5]{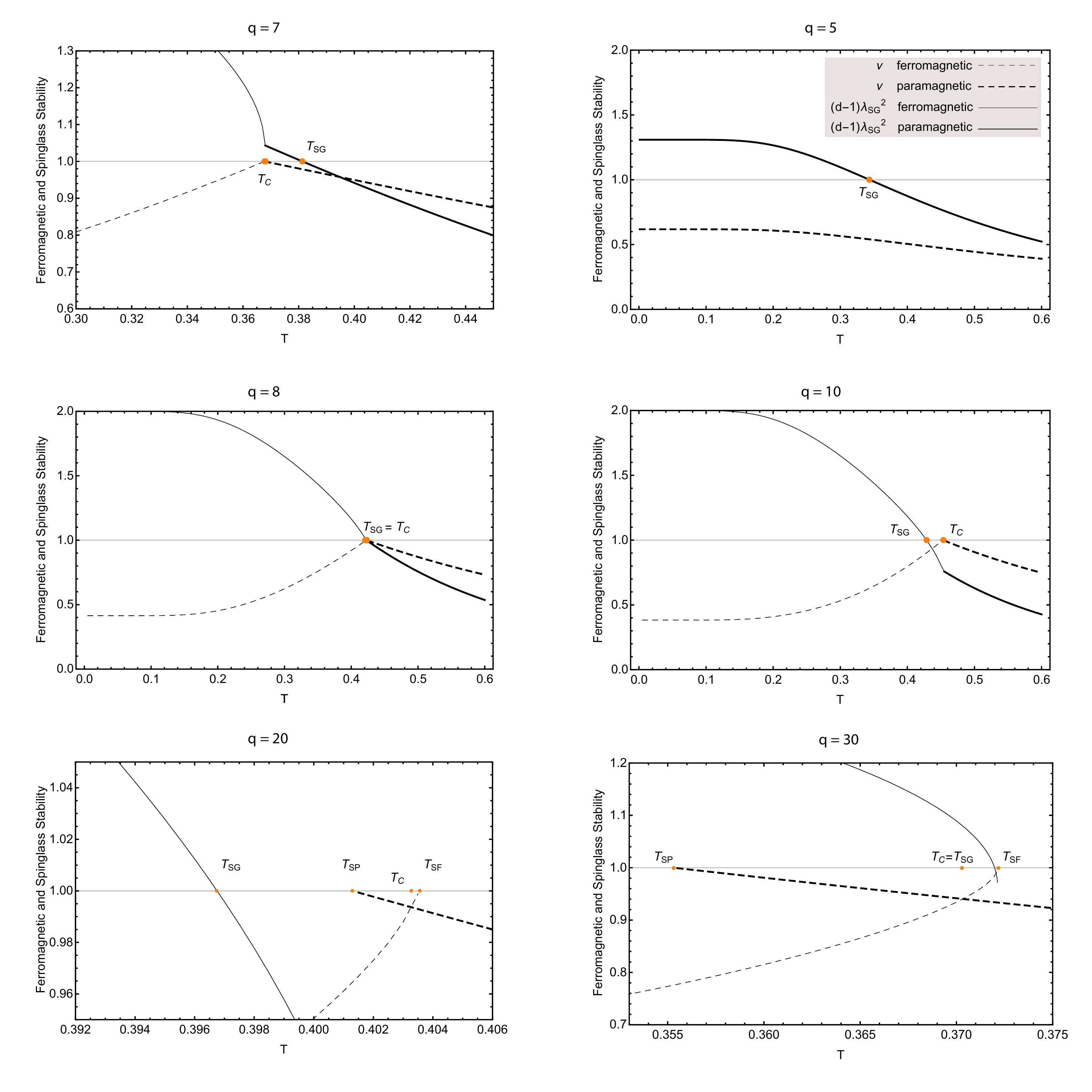}
\caption{Here we show the spin-glass stability
  condition as a function of the temperature for representative values
  of $q$ as discussed in the text. The full line represents the spin
  glass stability parameter of the thermodynamically stable replica
  symmetric solution (thick for paramagnet, thin for ferromagnet). The point where the lines cross one is the spin-glass temperature $T_{\mathrm{SG}}$. The
  dashed line represents the thermodynamic stability regions for the paramagnet
  (thick), and the ferromagnet (thin). In the last two cases ($q=20$ and $q=30$) the paramagnetic spinglass stability is smaller than the lower limit of the $y$ axis and does not appear in the figures. 
\label{fig:rs_transitions} }
\end{figure}

For $2 \leq q \leq 3$ circular coloring is equivalent to regular
coloring and the spin-glass temperature can be computed
\cite{ZdeborovaKrzakala07} from $T_{\mathrm{SG}}^{\rm col} = -1/\ln \left(  1-q/(1+\sqrt{d-1})\right)$.
For $3$-regular random graphs we get $T_{\mathrm{SG}} = 0.567$ when
$q=2$ and an always RS stable solution for $q=3$. When $4 \geq q
\geq6$ no ferromagnetic solution exists and the spin-glass transition
temperature is given by (\ref{eq:T_c}). We observe that for $q=7$ the
transition from RS to RSB happens in the paramagnetic phase before the
transition to the ferromagnetic state happens,
i.e. $T_{\mathrm{SG}}>T_{\mathrm{c}}$. At $q = 8$ the two transitions
coincide (compare equations (\ref{eq:T_c}) and (\ref{eq:T_SP})) and we have a tri-critical point: the system becomes
ferromagnetic at the same temperature where it loses its RS
stability. For $q\geq 8$ the ferromagnetic transition happens before
the RSB transition. Eventually, in the large $q$ limit the two
transitions coincide again and the ferromagnetic solution is never RS
stable and $T_{\mathrm{c}}=T_{\mathrm{SG}}$. For summary see figure
\ref{fig:rs_phase_diagram}, and figure \ref{fig:rs_transitions} for
examples of the instabilities.

\subsection{1RSB in the Energetic Zero Temperature Limit \label{sub:Energetic-Zero-Temperature-Limit-Results}}

In section \ref{sec:cavity_zero_temperature} we described the
1RSB energetic zero temperature limit. Using population dynamics for equations (\ref{eq:P(h)}) and (\ref{eq:Q(u)}) we obtain the replicated free energy $\Phi(y)$ and the energy $e(y)$ both as a function of $y$ in the Bethe approximation from (\ref{eq:replicated free energy}) and (\ref{eq:bethe_free_entropy}) respectively. The complexity as a function of the energy is then obtained from the Legendre transformation $-y\Phi(y)=-ye+\Sigma(e)$. The value of $e$ for which the concave
branch of $\Sigma(e)$ intersects the $e\textrm{-axis}$ yields the 1RSB
estimate for the ground state energy. 

The result of the energetic zero temperature cavity limit are very
different depending on whether the number of colors is even or
odd. Whereas for even number of colors, the results are exactly the
same as for $q=2$, for odd number
of colors the results can roughly be deduced from the replica
symmetric solution with a delay of the onset of the ferromagnetic
transition. 
 
Our result are obtained using the population dynamics, with an
initialization such that each initial field $\mathbf{h}$ in the
population is pointing into the direction of one color only,
i.e. $\mathbf{h} = \mathbf{e}_{\tau}$. And $\tau$ is drawn with probability equal to the replica symmetric estimate $P^{\rm{RS}}(\tau)$.

\subsubsection{Even Number of Colors \label{par:even_colors}}

In section \ref{sub:math} we argued that for even number of colors the
ground state energy of the circular coloring is the same as the ground
state of the $2$-coloring, i.e. the Viana-Bray model \cite{VianaBray85} for
which the energetic 1RSB results were studied in
\cite{MezardParisi03}. Indeed, assume $q=2z$ with $z\geq 2$ being an
integer. We identify $s=2z+1 \to 1$ and $s=2z \to 2$ for $s \in
\{1,\dots,q\}$ the number of contradictory edges in this $2$-coloring
must be equal or smaller than it was before. On the other hand any
$2$-coloring configuration is a valid configuration of $2z$-circular
coloring. This independence of $z$ also comes up from the energetic
zero temperature cavity limit solution, contrary to the RS approach. 

The 1RSB cavity method predicts
$E_{\mathrm{gs}}(q=2z,d)=E_{\mathrm{gs}}(q=2,c)$ with
$\Sigma\left[E(q=2z,d)\right]=\Sigma\left[E(q=2,d)\right]$. The SP-y
equations converge towards the same fixed point for all even $q$ when
initialized as described above. In this fixed point the population
consists of only two types of messages. The alternating warnings
$\mathbf{u}=(1,0,1,0,\dots)$ and $\mathbf{u}=(0,1,0,1,\dots)$ and the
trivial warning $\mathbf{u}=(0,0,\dots)$ with probabilities
$P\left[\mathbf{u}=(1,0,1,0,\dots)\right]=P\left[\mathbf{u}=(0,1,0,1,\dots)\right]=\eta/2$
and the zero warning $P\left[\mathbf{u}=(0,0,\dots)\right]=1-\eta$. The
corresponding ground state energies and zero energy complexities were
evaluated in details in previous works on the Viana-Bray model
\cite{MezardParisi03}. For instance for $d$-regular random graphs the
1RSB ground state energy is $e_{\rm gs}(d=3)=0.1138$, $e_{\rm
  gs}(d=4)=0.2635$, $e_{\rm gs}(d=5)=0.4124$ \cite{zdeborova2010conjecture}. 

\subsubsection{Odd Number Colors \label{par:odd_colors}}

\begin{figure}
\includegraphics[scale=0.5]{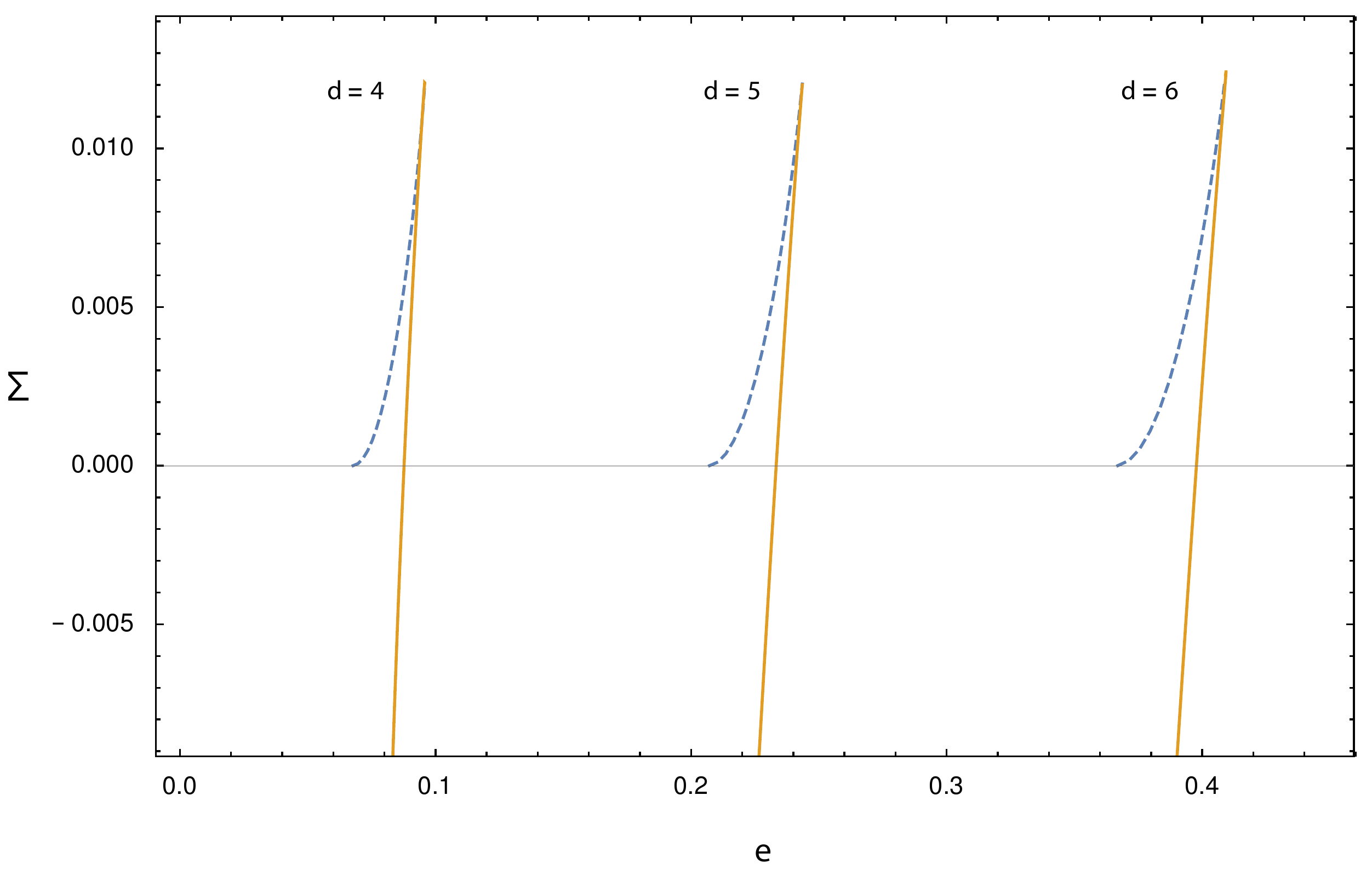}
\caption{
Some examples of the complexity curve as a function of the energy for $q=5$. The physical (concave) branches are depicted as full lines and the non-physical branches are dashed. The ground state energy can be extracted from the intersection point of the physical branch with the energy axis where $\Sigma(e) = 0$.}
\label{fig:Some-instances-of-SP-y}
\end{figure}

Contrary to the case of even colors which could be reduced to $q=2$
the population does not reduce to the simple subset of messages. In
fact, the  alternating warning \emph{cannot} exist if $q$ is an odd
number.  
Figure \ref{fig:Some-instances-of-SP-y} depicts some exemplary results,
table \ref{tab:tbl1} summarizes the ground state energies obtained for
different values of degree $d$ and number of colors $q$.

In particular we find that $E_{\mathrm{\rm gs}}>0$ for all considered
instances, except for $q=5,\ d=3$ the particular case for which we
only obtain the trivial solution $P\left[\mathbf{u} = (0,0,\dots)\right]$ when
employing the energetic zero temperature limit. These results agree
with what was found in previous rigorous investigations
cf. \cite{kostochka2001colorings,wanless2001regular,hatami2005random}. 

It can be expected from physical intuition, and was proven by M. Molloy (private
communication) that as the number of colors $q$ grows the ground state
energy for $q$ odd converges to the one for $q$ even. Our numerical
results are consistent with this. 

The RS investigation indicated the presence of a purely paramagnetic phase
for all temperatures when $q<q^{\mathrm{RS}}_{\mathrm{F}}$ and the
presence of a phase transition towards a ferromagnetic phase when
$q\geq q_{\mathrm{F}}^{\mathrm{RS}}$. In the framework of 1RSB we
monitor the magnetization by tracking an order parameter
$\bar{\Psi} = \mathrm{\int} \Psi P(\Psi) \mathrm{d}\Psi$. 
We observe a delay of the appearance of the ferromagnet at zero
temperature. In particular -- according to the 1RSB -- the ferromagnet
appears only at $q=11$ for $d=3$, $q=9$ for $d=4$ and $q=7$ for $d\geq
5$. Recalling the RS phase diagram for $d=3$ (figure
  \ref{fig:rs_phase_diagram}), one might wonder about the finite
  temperature behavior. The 1RSB solution for $q=7$ and $q=11$ are
  respectively of purely paramagnetic and ferromagnetic character in
  the whole range $T<T_{\textrm{SG}}$. However, anticipating the finite temperature investigation, for $q=9$ we find a reentrance of the 1RSB solution into a paramagnetic phase for $T < 0.125$.

\begin{table}
\begin{centering}
\begin{tabular}{|c|c||c|c|c|}
\hline 
$d$ & $q$ & RS phase& RSB phase  & $e_{\text{\rm gs}}$ \tabularnewline
\hline 
\hline 
$3$ & $5$ & para & para  & $0$ \tabularnewline
\hline 
$3$ & $7$ & ferro & para  & $0.030$ \tabularnewline
\hline 
$3$ & $9$ & ferro & para   & $0.064$\tabularnewline
\hline 
$3$ & $11$ & ferro & ferro & $0.052$  \tabularnewline
\hline 
$3$ & $13$ & ferro & ferro & $0.071$  \tabularnewline
\hline 
$3$ & $15$ & ferro & ferro & $0.076$  \tabularnewline
%\hline 
%$3$ & $17$ & ferro & ferro & $0.081$  \tabularnewline
\hline 
$3$ & $19$ & ferro & ferro & $0.084$  \tabularnewline
%\hline 
%$3$ & $23$ & ferro & ferro & $0.089$  \tabularnewline
%\hline 
%$3$ & $27$ & ferro & ferro & $0.093$  \tabularnewline
\hline 
$3$ & $29$ & ferro & ferro & $0.094$  \tabularnewline
\hline 
$3$ & $51$ & ferro & ferro & $0.100$  \tabularnewline
\hline 
%$3$ & $2n$ & -- & -- & para  & $-0.34$ & $0.111$ \tabularnewline %\ln(1+q/2\sqrt{2})^{-1}
%\hline 
\hline 
$4$ & $5$ & para & para & $0.088$ \tabularnewline
\hline 
$4$ & $7$ & ferro & para & $0.189$ \tabularnewline
\hline 
$4$ & $9$ & ferro & ferro & $0.155$  \tabularnewline
\hline 
$4$ & $11$ & ferro & ferro & $0.171$  \tabularnewline
\hline
%$4$ & $2n$ & -- & -- & para  & $-0.70$ & $0.263$ \tabularnewline %\ln(1+q/4)^{-1}
%\hline 
\hline 
$5$ & $5$ & ferro & para & $0.233$ \tabularnewline
\hline 
$5$ & $7$ & ferro & ferro & $0.325$ \tabularnewline
\hline 
$5$ & $9$ & ferro & ferro & $0.328$ \tabularnewline
\hline 
$5$ & $11$ & ferro & ferro & $0.337$ \tabularnewline
%\hline 
%$5$ & $2n$ & -- & -- & para  & $-1.04$ & $0.412$ \tabularnewline %\ln(1+q/2\sqrt{5})^{-1}
\hline 
\end{tabular}
\par\end{centering}
\caption{Results of the energetic 1RSB
  analysis for the circular coloring. The ground state energy
  $e_{\mathrm{\rm gs}}$ is obtained from the intersection point of the
  concave branch of $\Sigma(e)$ with the energy axis 
  (cf. figure \ref{fig:Some-instances-of-SP-y}). 
  We also provide which phase the
  zero temperature solution is found in the RS and 1RSB framework. The 1RSB
  zero temperature solution becomes ferromagnetic at $q=7$ for $d\geq 5$. }
\label{tab:tbl1}
\end{table}

\subsubsection{The case of $5$-circular coloring}

The very fact that the predictions obtained in the energetic zero temperature limit agree with previous rigorous results and the fact that even colors can always be reduced to $q=2$ is not trivial and provides some solid ground for expanding our investigations further to the particular case of $q=5,\ d=3$ for which the energetic zero temperature limit does not provide a useful non-trivial estimate.

\begin{figure}
\includegraphics[scale=0.5]{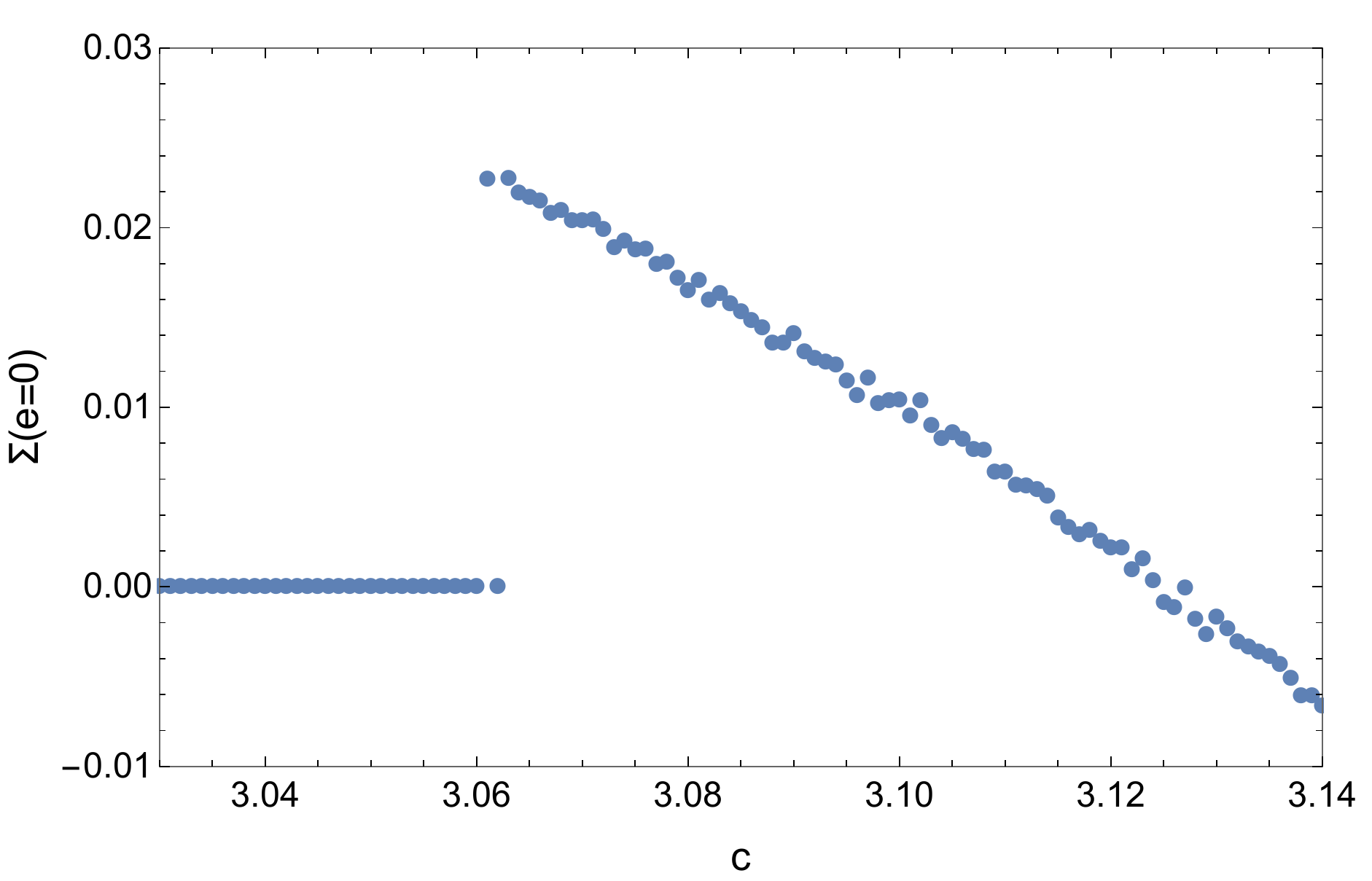}
\caption{Zero energy complexity against the fraction $r$ of degree $d=4$
  nodes in a graph with only degrees $3$ and $5$ for $5$-circular coloring. A noncontinuous jump happens around $c=3.06$ when the complexity suddenly becomes positive. Subsequently it is continuously decreasing and becomes negative around $c=3.13$.} \label{fig:complexity against ratio}
\end{figure}

The absence of non-trivial solution suggests that no constrained
fields are present, and thus $\Sigma(e=0)=0$. Consequently an
interesting follow up question is to ask what happens when the degree
is varied from $d=3$ to $d=4$. In the latter case constrained fields
are present and $\Sigma(e=0)\leq0$. In order to investigate the
intermediate regime where the average degree $3<c<4$ we work on the
ensemble of graphs with a fraction $1-r$ of degree $d=3$ nodes and
fraction $r$ of $d=4$. Tuning $r$ enables us to investigating the zero
energy complexity in the whole region $3\leq c \leq4$, where $ c = r
\cdot 4 + (1-r) \cdot 3 $. The survey propagation can be carried out
for each $r$ and the resulting $\Sigma(e=0,r)$ is depicted in figure
\ref{fig:complexity against ratio}. No non-trivial solution is found
and the complexity remains zero until a discontinuous jump happens
around $r=0.06$ where the complexity suddenly becomes positive and
constrained messages appear in the population. Subsequently the curve
is continuously decreasing and becomes negative around $r=0.13$ when
no zero energy solution can be found any longer. This suggests that
the local constraint density is sufficiently small for $r<0.13$ and
coloring is possible. However, no confined variables are present for
$r<0.06$ and to enable a quantitative statement for the case $q=5,\
d=3$ we must consult the entropic zero temperature limit.

We also considered the Erd\H{o}s-R\'{e}nyi graphs. Somewhat
  strangely we found that the zero energy complexity never becomes
  positive. Instead a solution with negative complexity appears at
  about $c \approx 2.27$. Recall that
  eq. (\ref{eq:eigenvalues_jacobian}) yields an average degree of
  $c_{\textrm{d}}=1.53$ for the spin glass stability transition
  point. We solved the
  corresponding survey propagation equations using population dynamics
  of populations with population sizes 10000. It cannot be excluded
  that what we observed was a finite population size effect, but it
  may also be that the usual 1RSB solution with confined variables
  that leads to the satisfiability threshold in the widely studied
  $K$-SAT and coloring problems is not sufficient to get a sensible
  estimate of the threshold for the $5$-circular coloring of the
  Erd\H{o}s-R\'{e}nyi random graphs

\subsection{1RSB in the Entropic Zero Temperature Limit\label{sub:Entropic-Zero-Temperature-Limit-Results}}

We briefly summarize the procedure that enables access to the thermodynamic quantities for the case $q=5,\ d=3$ before presenting the results. 
An advantage of working on regular graphs is the reduction of the required
computational effort necessary in the population dynamics. This is because the probability distribution over the messages in
Eq. (\ref{eq:1RSB recursion}) is identical for every edge
$P^{i\rightarrow j}(\mathbf{\psi})=P(\mathbf{\psi})$. 
For a given reweighting parameter $m$ the 1RSB equation (\ref{eq:1RSB recursion}) can be solved with the population dynamics technique. Solving it for every $m$ yields the thermodynamic quantities as function of the reweighting parameter $m$. 
These are $\Phi_{s}(m)$ which follows from (\ref{eq:replicated free energy}), $\Sigma(m)$ which is obtained from (\ref{eq:Legendre identities}) and $s(m)$ which can be computed with (\ref{eq:bethe_free_entropy}) and (\ref{eq:F=E-TS}) -- all in the limit $\beta\to\infty$ where $-\beta f = s$. In order to assign the right weights to each pure state in the equilibrium configuration (1RSB estimate), the reweighting parameter $m$ must be chosen accordingly. If $\Sigma(m=1)>0$ we are in the dynamic 1RSB phase and the 1RSB and RS solutions agree. Else if a non-trivial solution exists for $m=1$ and $\Sigma(m=1)<0$ we must choose the thermodynamic value of the reweighting parameter $m^{*}$ as the smallest positive non-zero value for which the complexity vanishes, i.e. $\Sigma(m^{*})=0$.

\begin{figure}
\includegraphics[scale=0.4]{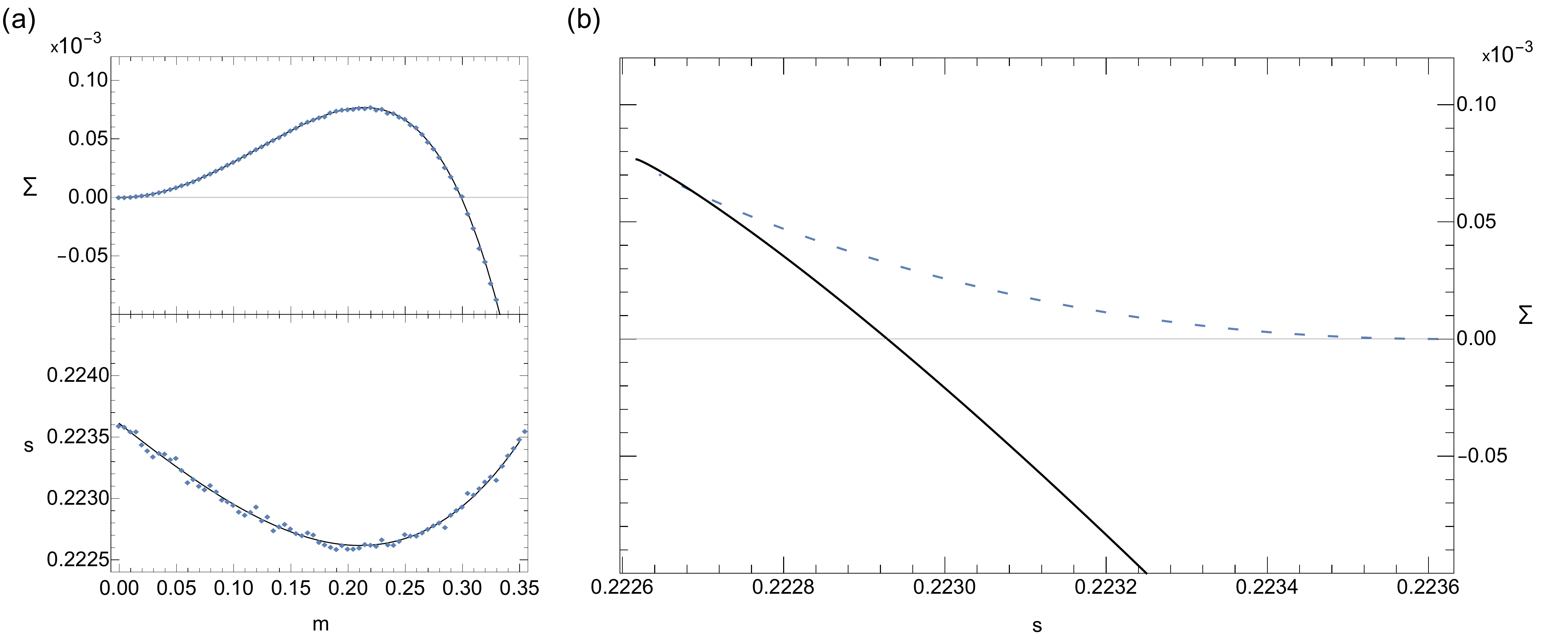}
\caption{Results obtained within the entropic zero temperature limit
  for $5$-circular coloring of $3$-regular random graphs. The two plots in (a) depict the complexity and entropy as a function of the reweighting parameter -- $\Sigma(m)$ and $s(m)$ respectively. Figure (b) presents the resulting complexity as a function of the entropy $\Sigma(s)$. The results were first fitted with a function of type $a+be^{-x}+ce^{-2x}+de^{-3x}+\dots$ and subsequently the change of variables from $\Sigma(m)$ to $\Sigma(s)$ was performed with the fit function. The resulting complexity function contains a physical branch (solid line) as well as an unphysical branch (dashed line). 
  The 1RSB estimate for the entropy of clusters of zero energy are obtained from the intersection of the physical branch of the complexity function with the entropy axis. \label{fig:1RSB-entropic-limit-figure}}
\end{figure}

Figure \ref{fig:1RSB-entropic-limit-figure} presents the results
obtained within the entropic zero temperature limit. In this limit the
energy vanishes $e=0$ and $\Sigma(s)$ counts the total number of
clusters of size $s$. The point where the physical (concave) part of
the complexity curve intersects with the entropy axis, i.e. where
$\Sigma(s)=0$, provides the 1RSB estimate for the entropy as the 
entropy of the entropically dominating clusters. 
Following the positive part of the physical branch of the complexity 
curve in figure \ref{fig:1RSB-entropic-limit-figure}b 
more and more (increasing complexity) subdominant 
clusters of smaller entropy appear. 
Eventually, at the cusp -- that is the point where the two branches of $\Sigma(s)$ meet in \ref{fig:1RSB-entropic-limit-figure}b where $s(m)$ is minimal in \ref{fig:1RSB-entropic-limit-figure}a -- the 1RSB solution loses its validity (it becomes unstable towards more levels of RSB before the cusp is reached). 

The thermodynamic value of the reweighting parameter was found to be
$m^{*}=0.299$ with an 1RSB entropy estimate of the entropy
$s_{\mathrm{1RSB}}=0.223$ (to be compared with the replica symmetric
estimate of the entropy $s_{\rm RS}=0.235$). Finding $m^{*}(T=0)\neq1$
(or more precisely the fact that $\Sigma(m=1)<0$) impose that the
solution space of the problem is behind the condensation transition
\cite{KrzakalaMontanari06}. This 1RSB result is a very strong indication
towards $5$-circular colorability of $3$-regular random graphs. 
In the next section we will show that the 1RSB approach is actually not
stable, and further steps of replica symmetry breaking are
necessary. However, these effects usually change the value of the
1RSB ground state entropy only very little, so we conjecture that the
colorability remains.

\subsection{Finite Temperature and 1RSB Stability}

In this section the $m,\ T$ phase diagram for the $5$-circular
coloring of $3$-regular random graphs is investigated in order to
study the stability of the 1RSB solution towards further steps of
symmetry breaking within the whole regime $T<T_{\mathrm{SG}}$, where
$T_{\mathrm{SG}}$ indicates the temperature at which the RS estimate
becomes unstable. With $T_{\mathrm{SG}}=0.344$ for the five circular
coloring of three regular graphs. 

\begin{figure}
\includegraphics[scale=0.5]{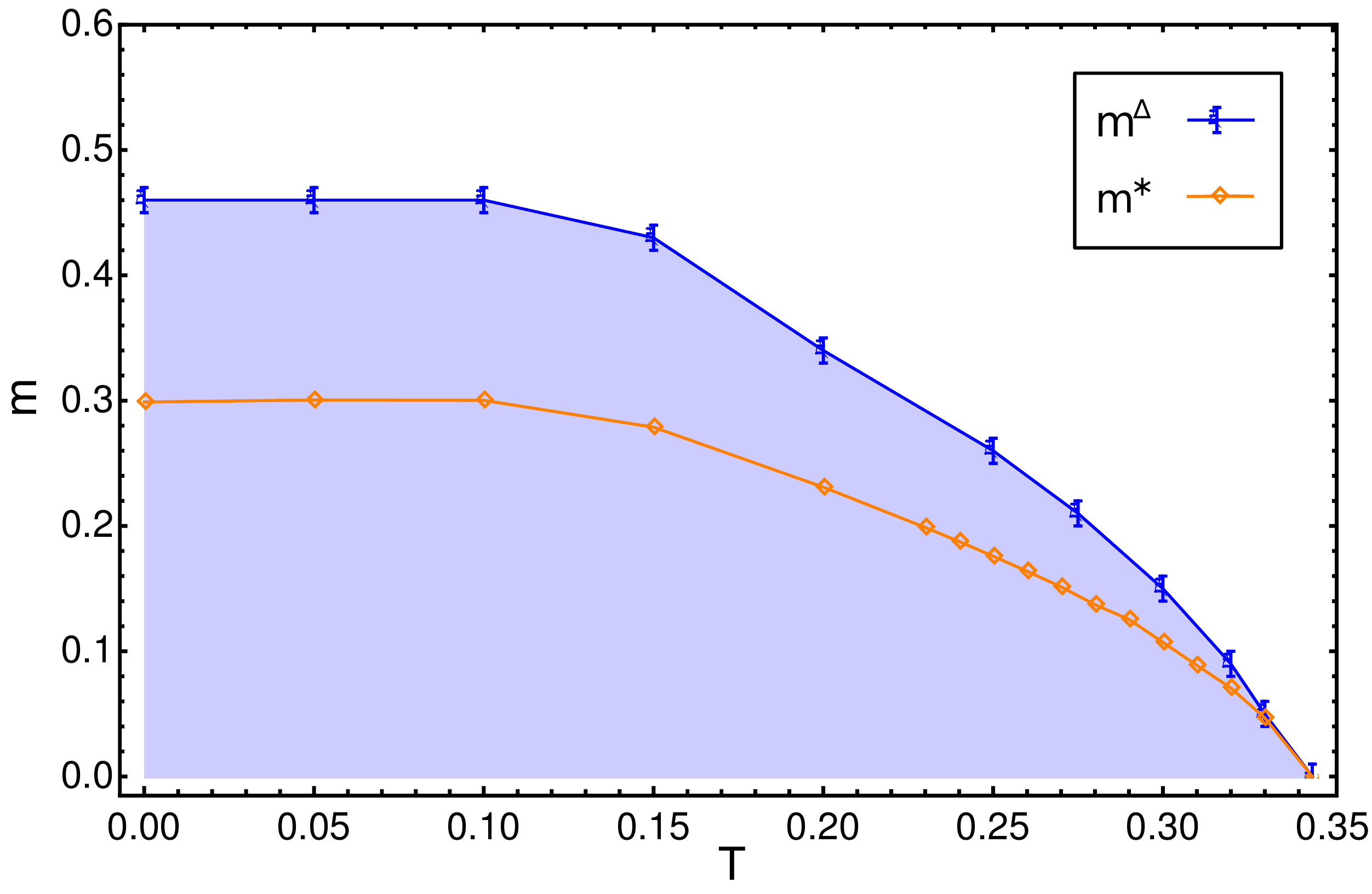}
\caption{The $m-T$ phase diagram for $T<T_{\mathrm{SG}}=0.344$ for the
  $5$-circular coloring of $3$-regular random graphs. The blue squares mark the values $m^{\Delta}$ at which the 1RSB solution becomes unstable. The whole shaded area below the $m^{\Delta}$-line is unstable towards further steps of RSB. The orange diamonds denote the thermodynamic values of the reweighting parameter $m^{*}$, extracted as $\Sigma(m^{*})=0$. For all relevant temperatures $m^{*} < m^{\Delta}$ holds and hence the 1RSB solution is unstable and further steps of replica symmetry breaking are necessary. \label{fig:Stability Results}}
\end{figure}

The stability of 1RSB towards more levels of RSB is related to the
evolution of an infinitesimal perturbation in the messages, $Q(t)$, as
we described in Section \ref{sub:1RSB-framework}. The solution is considered stable towards further steps of RSB if $\lim_{t\to\infty}Q(t)\to 0$. The results are presented in figure \ref{fig:Stability Results}, where the thermodynamic value of the reweighting parameter, $m^{*}$, as well as the value for which the instability sets in, i.e. $m^{\Delta}$, are plotted as functions of the temperature. For a given temperature $m^{*}(T)$ was extracted as the point where $\Sigma\left[m^{*}(T)\right]=0$ and $m^{\Delta}(T)$ as the point where $Q(t)\geq Q(0)$ for large enough $t$. In the numerical tests $t=1000$ was used. 

As depicted in Fig. \ref{fig:Stability Results} the 1RSB solution
lays in the unstable (shaded) region, i.e. $m^{*}<m^{\Delta}$, for all
$T<T_{\mathrm{SG}}$ and hence the 1RSB solution is unstable for all
relevant temperatures. Consequently the 1RSB solution is not
sufficient in this case and further steps of replica symmetry breaking
are necessary. In the cases where the 1RSB framework fails it is
widely believed that the FRSB framework is necessary for a
correct treatment. Interestingly the
$q=5,\ c=3$ case therefore appears to be the unique case of a CSP with
exactly known degenerated zero ground state energy and FRSB structure
of that we are aware. On the one hand the FRSB structure makes it
unlikely that rigorous results will be obtained via the cavity
method. On the other hand it makes it a very interesting instance to
be studied in terms of algorithmic consequences. It has previously
been argued that a FRSB structure of the solution space is related to
marginal stability and absence of basins that would trap physical dynamics
for exponential time \cite{CugliandoloKurchan1994}. As a consequence
simple search algorithms, such as simulated annealing, are expected to
converge towards an optimal solution which we shall confirm in the
next section. Therefore it might be plausible to design an algorithm that
would provably reach the ground state in this problem. This is an
interesting direction for future work. Besides the algorithmic consequences it also is an interesting instance to obtain further insights on the nature of FRSB as it is the first case for which the FRSB framework is necessary although the correct ground state energy is known.

\subsection{Algorithmic Consequences \label{sub:Algorithmic-Consequences}}

\begin{figure}
	\includegraphics[width=0.7\textwidth]{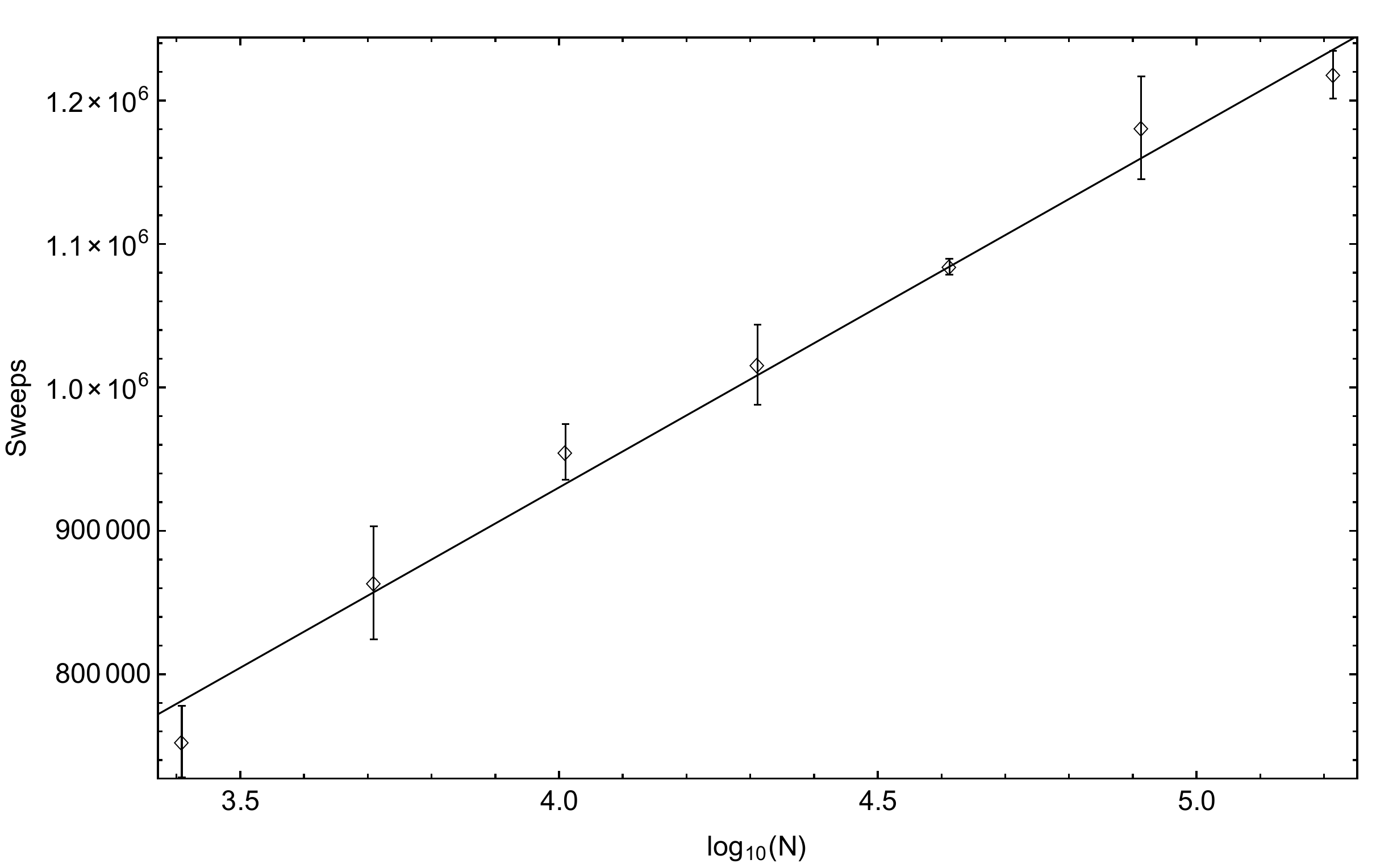}
	\caption{We plot the number of sweeps necessary to find a
          proper $5$-circular coloring or a random $3$-regular graph
          as a function of the size of the system $N$ for exponential annealing rate $\delta T = 10^{-6}$.}
	\label{fig:SA_exp}
\end{figure}

In order to investigate the performance of simulated annealing (SA)
for $5$-circular coloring of $3$-regular random graphs, we created
3-regular random graphs by randomly linking $N$ nodes such that no
self-loops, double edges or triangles are present (we simply dismiss
the graphs containing triangles). We then assign colors to the $N$ nodes uniformly at random and SA is performed. For each instance we considered different system sizes $N$ and ran SA with different annealing rates $\delta T$ for each of them. 
We have tried different annealing schedules and found that an
exponential variation with $T \to \frac{T}{1+\delta T}$ is very
efficient. We have implemented the schedule and present the results in
figure \ref{fig:SA_exp}. We plot the number of sweeps necessary to
find a proper coloring as a function of the systems size $N$ for a fixed
annealing rate $\delta T$. Each sweep takes $N$ steps and the
necessary number of sweeps is in good agreement with a logarithmic fit
(or smaller). Therefore the total number of iterations appears to be close to $O (N \log N)$ which makes SA equipped with the exponential schedule a very efficient choice to find proper colorings for five-circular coloring of three regular graphs.
We investigated belief propagation initialized in the solutions obtained with
simulated annealing, and after sufficient number of iterations the BP
behaves exactly in the same way it does when initialized randomly. 

In the previous section we saw that the problem has zero ground state
energy, that no confined variables are present \emph{and} that it
likely features a FRSB structure. The very fact that simulated
annealing as a simple search algorithm is able to find optimal
solutions supports the picture of a saddle point dominated landscape
of solutions that may account for the algorithmic easiness of five
circular coloring on $3$-regular random graphs. 

The success of SA for $q=5$ and $d=3$ inspired us to approach the problem with
an even simpler greedy strategy, that did not work. We find useful reporting 
that result here anyhow. The greedy strategy successively assigns constraints 
to the nodes of the graph and hence attempts to obtain a coloring. Given an 
unconstrained graph, we start by picking a random node and constraining it to 
color $s$. Doing so, we constrain all direct neighbors to $\{s-1,s+1\}$, all 
second neighbors to $\{s,s-2,s+2\}$ and all third neighbors to 
$\{s-1,s-2,s+1,s+2\}$ -- all $\textrm{mod}\,5$. After this step we pick one of 
the maximally constrained nodes at random, we assign it one of the permitted 
colors at random and subsequently update the constraints of its neighborhood. 
Then we repeat picking a maximally constrained node and proceed as before. We 
observe that the probability of success is very small for small graphs and 
decreases further with the size of the graphs.

\section{Conclusion}

Motivated by Ne\v{s}et\v{r}il's conjecture on $5$-circular
colorability of every high girth sub-cubic graph, this paper studied
the circular colorability of random regular graphs. The problem is
approached with the statistical physics based cavity method. After
analyzing the replica symmetric solution, the one step replica
symmetry breaking and its stability was investigated. In this framework the entropic as well as the energetic zero temperature limit were performed in order to study the ground state energies.

Within the replica symmetric Ansatz, we found that the $q-T$ phase
diagram splits into three regions. When the number of colors
$q<q_{\mathrm{F}}^{\mathrm{RS}}$ we are in the purely paramagnetic
region and the only solution is the paramagnetic fixed point
$\Psi^{i\rightarrow j}_{s_i}=1/q$ for all $T$. Subsequently,
when $q\geq q^{\mathrm{RS}}_{\mathrm{F}}$ a phase transition separates
the paramagnetic high temperature region from a ferromagnetic low
temperature region. In the 1RSB approach we confirmed the presence of
a purely paramagnetic phase for low values of $q$ and the presence of
a low temperature ferromagnetic phase for higher and odd values of $q$. We find that the ferromagnetic solution appears later than predicted within the RS approach, i.e. $q^{\mathrm{RSB}}_{\mathrm{F}} > q^{\mathrm{RS}}_{\mathrm{F}}$.

Within the energetic 1RSB zero temperature limit the structure of the
warning propagation equations is significantly more involved compared
to previously studied problems, which makes the analytical treatment
rather cumbersome. In previously considered CSPs the variables could
either be trivial or point into one of the $q$ possible directions
$\mathbf{e}_{\tau}$. That is no longer the case in circular coloring
and the relevant warnings are from a larger domain. Such cases were
not much studied and in fact we did not manage to write the resulting
the survey propagation equations in a general explicit form. However,
a numerical circumvention is possible and we studied several cases of
regular random graphs with population dynamics. We found that the
energetic zero temperature limit results agree with previous rigorous results. Interestingly the 1RSB investigation reproduces the fact that circular coloring with an even number of colors can always be reduced to $q = 2$. The five circular coloring for three regular random graphs appears to be the particular case for which no non-trivial results is obtained in the energetic zero temperature limit, suggesting the absence of confined variables (in analogy to frozen variables). 

In order to resolve the zero temperature limit for $q=5$ on
$3$-regular random graphs we considered the entropic zero limit and
revealed that a typical instance is indeed $5$-circular colorable --
providing evidence for Ne\v{s}et\v{r}il's Pentagon conjecture. The
problem was found to fall into the condensed phase with a
thermodynamic value of the reweighting parameter of $m^{*}=0.299$ and
an 1RSB entropy estimate of $s_{\mathrm{1RSB}}=0.223$. For further
analysis of the stability of the 1RSB solution the finite temperature
phase space was studied and the 1RSB solution was found to be unstable
towards two-step replica symmetry breaking, implying likely that the problem requires the full replica symmetry breaking framework. 

To the best of the authors knowledge $5$-circular coloring of
$3$-regular random graphs is hence the first instance of a
\emph{satisfiable} combinatorial problem with degenerated and \emph{precisely known ground state} that requires FRSB. Thereby making it a very interesting problem to be studied in the context of diluted models on its own. The fact that the ground state energy is known does also give rise to the hope for simplifications in rigorous investigations along the same lines.

Finally the algorithmic consequences were examined by applying
simulated annealing to typical instances of the circular coloring
problem. We conclude that an exponential annealing schedule is very
efficient to find proper coloring assignments. The numerical
investigations suggest that SA equipped with the exponential schedule
scales like $O(N \log N)$. Remarkably although the statistical
description of the space of proper colorings seems challengingly hard
for the case of 5-circular coloring of 3-regular random graphs, the
problem is algorithmically easy. This suggests that a constructive
(algorithmic) proof Ne\v{s}et\v{r}il's conjecture might be much easier
than a non-constructive probabilistic or combinatorial proof. Further numerical investigation of this system might also shed light on the very nature of FRSB in diluted systems by taking advantage of the fact that in the present case we know the exact ground state energy.

\section{Acknowledgment }

We would like to thank Jaroslav Ne\v{s}et\v{r}il for suggesting this
interesting problem to us, and to Cris Moore for suggesting to look at
the greedy algorithm. This work is supported by the "IDI 2015" project funded by the IDEX Paris-Saclay, ANR-11-IDEX-0003-02.

\bibliographystyle{apsrev4-1}
\bibliography{refs.bib}

%\appendix

\end{document}